\documentclass[a4paper, 11pt]{article}
\usepackage{graphicx} 
\usepackage[english]{babel}
\usepackage[toc,page]{appendix}
\usepackage{color}
\usepackage[utf8]{inputenc}
\usepackage[T1]{fontenc}
\usepackage{amsmath, amsthm, amssymb}
\usepackage{mathtools}
\usepackage{mathrsfs}
\usepackage{bbm}
\usepackage{enumitem}
\usepackage{hyperref}
\usepackage{geometry}
\usepackage{dsfont}
\usepackage{csquotes}
\usepackage[dvipsnames]{xcolor}

\usepackage{import}
\usepackage[doi=false,isbn=false, url=false, date=year, backend=biber,maxbibnames=99
]{biblatex} 
\renewbibmacro*{journal+issuetitle}{%
  \usebibmacro{journal}%
  \setunit*{\addspace}%
  \iffieldundef{series}
    {}
    {\newunit
     \printfield{series}%
     \setunit{\addspace}}%
  \usebibmacro{volume+number+eid}%
  \setunit{\bibpagespunct}%
  \printfield{pages}%
  \setunit{\addspace}%
  \usebibmacro{issue+date}%
  \setunit{\addcolon\space}%
  \usebibmacro{issue}%
  \newunit}

\renewbibmacro*{note+pages}{%
  \printfield{note}%
  \newunit}

\theoremstyle{plain}
\newtheorem{theorem}{Theorem}[section]
\newtheorem{lemma}[theorem]{Lemma}
\newtheorem{proposition}[theorem]{Proposition}
\newtheorem{corollary}[theorem]{Corollary}

\theoremstyle{definition}
\newtheorem{definition}[theorem]{Definition}

\theoremstyle{remark}
\newtheorem{remark}[theorem]{Remark}

\DeclareMathOperator{\aut}{Aut}
\DeclareMathOperator{\Lker}{ker^L}

\newcommand{\E}{\mathbb{E}}
\newcommand{\tr}{\mathrm{tr}}
\newcommand{\e}{\mathrm{e}}
\newcommand{\diam}{\mathrm{diam}}
\newcommand{\dd}{\mathrm{d}}
\renewcommand{\d}{\mathrm{d}}
\renewcommand{\i}{\mathrm{i}}
\renewcommand{\Phi}{\varPhi}
\renewcommand{\Psi}{\varPsi}
\renewcommand{\Sigma}{\varSigma}

\newcommand{\g}{\gamma}

\newcommand{\mI}{\mathcal{I}}

\newcommand{\w}{\omega}
\newcommand{\N}{\mathbb{N}}
\newcommand{\Z}{\mathbb{Z}}
\newcommand{\C}{\mathbb{C}}

\newcommand{\mA}{\mathcal{A}}
\newcommand{\R}{\mathbb{R}}
\newcommand{\eps}{\varepsilon}

\newcommand{\OD}[1][]{   ^{\mathrm{OD}#1}   }
\newcommand{\D}[1][]{   ^{\mathrm{D}#1}   }
\newcommand{\mL}{\mathcal{L}}

\usepackage{mathtools}

\DeclarePairedDelimiter\norm{\lVert}{\rVert}

\DeclarePairedDelimiter\br{\lparen}{\rparen}

\renewcommand{\Dot}[1]{#1'}

\newcommand\numberthis{\addtocounter{equation}{1}\tag{\theequation}}

\usepackage{parskip}

\title{Automorphic equivalence within  gapped phases of infinitely extended   fermion systems}
\author{Lennart Becker%
\texorpdfstring{\footnote{\parbox[t]{.7\textwidth}{
                \foreignlanguage{ngerman}{Fachbereich Mathematik,  Universität Tübingen,\\
                Auf~der~Morgenstelle~10, 72076~Tübingen,} Germany
            }
        }
    }{}%
\and Stefan Teufel%
\texorpdfstring{%
        \footnotemark[1]
    }{}%
\and Marius Wesle%
\texorpdfstring{%
        \footnotemark[1]
    }{}%
}
\date{\today}

\begin{document}

\maketitle

\begin{abstract}
We prove automorphic equivalence within gapped phases of infinitely extended lattice fermion systems (as well as spin systems) with super-polynomially decaying interactions.
As a simple application, we prove a version of Goldstone's theorem for such systems: if an infinite volume interaction is invariant under a continuous symmetry, then any gapped ground state is also invariant under that symmetry.
\end{abstract}

\section{Introduction}

The so-called quasi-adiabatic evolution connects continuous families of gapped ground states in finite quantum spin systems by a cocycle of automorphism   generated by an explicit quasi-local interaction. It was first introduced by Hastings in \cite{hastings2007quasi} and formalized more fully in  \cite{Bachmann2011}. It has since become a standard tool for analyzing gapped many-body systems with finite-range or exponentially decaying interactions.
 Moon and Ogata \cite{moon2019automorphic} constructed the quasi-adiabatic evolution automorphisms also in infinite volume for spin systems with finite-range interactions, assuming only a gap for the infinite system (in the sense of a locally unique gapped ground state, see Definition~\ref{def:gap}). This is relevant in the context of topological materials, for example, which have no gap in finite volume due to edge states.

In the following we generalize the  result of \cite{moon2019automorphic} to systems with interactions that decay only super-polynomially, as well as to lattice fermion systems. Both generalizations require additional non-trivial modifications of and additions to the original proof strategy. 
In particular, they rely on recent  Lieb-Robinson bounds with algebraic light-cones for fermion systems with polynomially decaying interactions \cite{TeufelWessel25}. We expect our results to be useful in extending existing   applications of the quasi-adiabatic evolution to infinitely extended lattice systems with super-polynomially decaying interactions. For instance, in a companion paper, we generalize the many-body adiabatic theorem \cite{BDF2018,MT2019,T2020,Henheik_Teufel_2022} to bulk-gapped fermion  systems with super-polynomially decaying interactions, based on the results of the present paper.

One new application of our main result on quasi-adiabatic evolution, which we discuss in this paper, is a new, simple proof of the   Goldstone theorem. Assume that a spin or lattice fermion system has an interaction $H$ that decays super-polynomially and is invariant under a continuous symmetry, i.e.\ it commutes with the generator of a one-parameter group of automorphisms of the quasi-local algebra. If this generator is also given by a super-polynomially decaying interaction, then any gapped ground state of $H$ is also invariant under that symmetry. It should be noted that Goldstone's theorem is typically expressed as the contraposition of this statement: no ground state that breaks a continuous symmetry can be gapped. Although Goldstone's theorem originated in quantum field theory, its relevance to quantum statistical mechanics has also long been acknowledged, see for example the recent review \cite{frohlich2023phase} and references therein.  For infinitely extended quantum spin systems with finite range interactions it was proved in \cite{wreszinski1987charges} under an additional ergodicity assumption. A different proof for $U(1)$-symmetry and exponentially decaying interactions was sketched in \cite{kapustin2020hall}.

Let us explain our main result concerning automorphic equivalence within gapped phases (see Theorem~\ref{Theorem: main theorem})  in some more detail. We consider  fermions on the lattice $\Z^d$ with super-polynomially decaying interactions. However, all statements hold by basically the same or simpler proofs also for spin systems on $\Z^d$. One novelty compared to \cite{moon2019automorphic} is that we work in the fixed Fr\'echet spaces $\mA_\infty$ resp.\ $\mathcal{P}_\infty$  of super-polynomially decaying operators resp.\ interactions instead of using nested Banach spaces defined by explicit decay functions as in \cite{moon2019automorphic}. This makes some of the technical steps simpler or at least more transparent. 
Our main result  is the following:
Let $H: I\to \mathcal{P}_\infty$, $s\mapsto H_s$ be a   differentiable family of interactions and assume that for all $s\in I$  the interaction $H_s$ has  a locally unique gapped ground $\omega_s$ (see Definition~\ref{def:gap}) such that  $s\mapsto \omega_s(A)$ is differentiable for all $A\in \mA_\infty$ and $ \Dot{\omega}_s:\mA_\infty \to \C$ is continuous for all $s\in I$. Then the  cocycle of automorphisms $(\alpha_{u,v})_{(u,v)\in I^2}$ generated by the family of interactions $(-\mathcal{I}(  \Dot{H}_s))_{s\in I} \subset \mathcal{P}_\infty$ satisfies
\begin{equation}\label{eq:autoeqintro}
\omega_t = \omega_s\circ \alpha_{s,t} \qquad \mbox{for all $s,t\in I$.}
\end{equation}
Here $\mathcal{I}$ is the quasi-local inverse of the Liouvillian, meaning, in particular, that
\begin{equation}
\label{eq:invliouintro}
    \i[H_s, -\mathcal{I}(  \Dot{H}_s)] \eqcolon (  \Dot{H}_s)\OD[s]
\end{equation}
is the off-diagonal part  of $ \Dot{H}_s$ with respect to $\w_s$.

The Goldstone theorem (see Theorem~\ref{thm: invariance of locally unique gs}) now follows as a simple corollary: Assume that $H$ has a gapped ground state $\omega_0$ and that $\R\ni s\mapsto \beta_s$ is a one-parameter family of automorphisms generated by an interaction in $\mathcal{P}_\infty$ such that $\mathcal{L}_{H}\beta_s =\beta_s \mathcal{L}_{H}$, i.e.\ a family of locally generated automorphisms that leaves $H$ invariant (a ``continuous symmetry of $H$'').  It is then straightforward to check that $s\mapsto \omega_s := \omega_0 \circ \beta_s$ is a family of gapped ground states for $H_s\equiv H$. But our main result then implies that  $\omega_s = \omega_0\circ \alpha_{0,s}$, where the cocycle $\alpha_{0,s}$ is generated by $-\mathcal{I}(\Dot{H}_s)$.  Since $H_s\equiv H$ is constant, $-\mathcal{I}(\Dot{H}_s)= 0$ and thus $\alpha_{0,s}= \mathrm{id}$   and   $\omega_s=\omega_0$ for all $s\in\R$. 
 \medskip

\begin{remark}
To better appreciate the result on automorphic equivalence, let us compare it to the corresponding statement for families of self-adjoint operators on Hilbert spaces. Let  $H: I\to \mathcal{L}(\mathcal{H})$, $s\mapsto H_s$, be a differentiable family of self-adjoint bounded operators on a Hilbert space $\mathcal{H}$, and assume that for all $s\in I$ the infimum $E_s$ of the spectrum of $H_s$ is separated by a gap from the rest of the spectrum. Let $  P_s$ denote the corresponding ground state projection  of $ H_s$. Then, by a standard argument, it follows that $s\mapsto P_s$ is also differentiable. Next, differentiating $P_s^2=P_s$ with respect to $s$ implies that $\Dot{P}_s$ is off-diagonal, i.e.\ $\Dot{P}_s= P_s \Dot{P}_s P_s^\perp +P_s^\perp \Dot{P}_s P_s$, and thus that $P_s$ solves the linear equation
\[
    \i   \Dot{P}_s = [K_s,P_s]
\]
with the generator of parallel transport given by $K_s = \i [\Dot{P}_s,  P_s]$.
As a consequence, the cocycle of unitaries $U_{v,u}$ generated by $K_s$,
\[
\i \partial_v U_{v,u} = K_v U_{v,u}
\]
intertwines the different ground state projections,
\begin{equation}\label{eq:partransHamilt}
P_t = U_{t,s}\, P_s\, U_{s,t}\qquad \mbox{for all $s,t\in I$,}
\end{equation}
which is the Hilbert space version of \eqref{eq:autoeqintro}. And indeed, 
\begin{eqnarray*}
[H_s, K_s] &=&\i [ H_s,  [\Dot{P}_s,  P_s]] \;=\; -\i( E_s (P_s  \Dot{P}_s +   \Dot{P}_s P_s) - ( H_s   \Dot{P}_s P_s + P_s  \Dot{P}_s  H_s))\\
&=& 
-\i ( E_s  \Dot{P}_s + P_s^\perp H_s   \Dot{(P_s^\perp)} P_s +  P_s   \Dot{(P_s^\perp)}  H_s P_s^\perp) 
\;=\; -\i ( P_s \Dot{ H}_s P_s^\perp + P_s^\perp \Dot{ H}_s P_s)\\
&=:&-\i (  \Dot{ H}_s)\OD[s]\,,
\end{eqnarray*}
the analogue of \eqref{eq:invliouintro}.
While \eqref{eq:partransHamilt} follows from a straightforward computation, the proof of the corresponding infinite-volume result, \eqref{eq:autoeqintro}, is more subtle. This is because the GNS representations of $\omega_s$ for different $s$ need not be unitarily equivalent, so one can not just follow the same argument in some GNS representation. Let us comment on the differences in some more detail:  
Firstly, the differentiability of $s\mapsto \omega_s$ is now an assumption, rather than a consequence of the gap condition and the differentiability of $s\mapsto H_s$. 
Secondly, the notion of the off-diagonal part of an operator or an interaction with respect to $\omega_s$ is more subtle than simply taking the off-diagonal part in  a block decomposition with respect to the orthogonal projection $P_s$. Third and most difficult,  the parallel transport condition
$\Dot{\omega}_s(A) = \Dot{\omega}_s(A\OD[s]) $   no longer has a straightforward proof, unlike    the analogue  $\mathrm{tr}(\Dot{P}_s A) = \mathrm{tr}(\Dot{P}_s A\OD[s])$  in the Hilbert space case, which just follows from $\Dot{P}_s = (\Dot{P}_s)\OD[s]$, which in turn follows from differentiating $P_s^2=P_s$, as we saw above. 
\end{remark}
\medskip

\begin{remark}
Let us  briefly comment also on the required changes for the generalization to lattice fermions. While one could just restrict to even observables and follow the arguments of \cite{moon2019automorphic}, this would result in Equation~\eqref{eq:autoeqintro} for even observables only. 
To obtain the full automorphic equivalence, one would have to assume in addition that the states $\w_s$ are even for all $s$.
By avoiding this assumption in our argument, we are actually able to conclude a posteriori through Corollary~\ref{cor: gauge-invariance} that if the Hamiltonians $H_s$ are gauge-invariant, then the states $\w_s$ are always gauge-invariant (and in particular even).

 The proof of  \cite[Lemma 4.1]{moon2019automorphic}, the statement analogous to the continuity of the time-evolution $\e^{\i t \mL_{H_s}}$ as a map from  $\mA_\infty$  to $\mA_\infty$, relies on evenness through the use of   \cite[Proposition A.1]{moon2019automorphic}. This standard proposition states that for a quasi-local observable $A \in \mA$ and a finite set $M \subseteq \Z^d$ that if for all local observables $B$ supported outside of $M$ it holds that $\lVert [A,B] \rVert \leq \eps   \lVert B \rVert$, then one also has $\lVert A- \E_M A \rVert \leq 2 \eps$, where $\E$ is the conditional expectation. In the fermionic case it is not sufficient to bound the commutator only for even $B$, meaning that useful bounds can only be obtained in this way for even observables $A$. 

We circumvent this issue by bounding the expression $\Vert \e^{\i t\mL_{H_s}}A - \E_M\e^{\i t\mL_{H_s}} A \rVert$ in another way. We follow the usual calculation that is used to show  existence of the infinite-volume dynamics (see for example \cite{Bru2017}), which uses Lieb--Robinson bounds to show that suitably defined finite-volume approximations of the dynamics on increasing boxes form a Cauchy sequence. The calculation is then modified through the use of the decay properties of $H_s \in \mathcal{P}_\infty$ and $A\in \mA_\infty$, expressed via the fermionic conditional expectation, to obtain a super-polynomial rate of convergence. This then results in $\e^{\i t \mL_{H_s}} A \in \mA_\infty$ also for non-even observables $A$. It is also important to note that for our purposes (defining the infinite-volume spectral flow) the bound on $\Vert \e^{\i t\mL_{H_s}}A - \E_M\e^{\i t\mL_{H_s}} A \rVert$ may not grow too quickly in $t$. For finite-range interactions one achieves this through the standard linear light-cone Lieb--Robinson bounds, which are not available for our class of interactions. Instead, we obtain Lieb-Robinsin bounds with an algebraic light-cone, starting from the  Lieb--Robinson bounds in \cite{TeufelWessel25}. See Lemma \ref{lem: cocycles} and Appendix \ref{sec: Lieb-Robinson bounds} for details. 
\end{remark}

\medskip
Our paper is structured as follows. In Section~\ref{sec:setup} we introduce the basic mathematical setup and notation. In Section~\ref{sec:results} we formulate our main results and give the core arguments of the proof deferring proofs of technical tools and lemmas to the next section resp.\ to several appendices.
In Section~\ref{sec:partrans} we prove the key ingredient to  automorphic equivalence, namely the parallel transport condition $ \Dot{\omega}_s(A) = \Dot{\omega}_s(A\OD[s]) $ mentioned above. The three appendices then contain proofs of further technical tools and lemmas.

\section{Mathematical Setup}\label{sec:setup}

The anti-symmetric  (or fermionic) Fock space over the lattice $\Z^d$ is  
\begin{align*}
   \mathcal{F}(\Z^d,\C^n) \coloneq \bigoplus_{N=0}^{\infty}\ell^2(\Z^d,\C^n)^{\wedge N}.
\end{align*}
We use $a^*_{x,i}$ and $a_{x,i}$ for $x\in \Z^d$, $i\in \{1,\dots,n\}$, to denote the fermionic creation and annihilation operators associated to the standard basis of $\ell^2(\Z^d,\C^n)$ and recall that they satisfy the canonical anti-commutation relations (CAR).
The number operator at site $x\in\Z^d$ is defined by
\begin{align*}
    n_x \coloneq \sum_{i=1}^n a^*_{x,i}a_{x,i}\,.
\end{align*}
The algebra of all bounded operators on $\mathcal{F}(\Z^d,\C^n)$ is denoted by $\mathcal{B}(\mathcal{F}(\Z^d,\C^n))$. 
For each $M\subseteq \Z^d$ let $\mA_M$ be the C$^*$-subalgebra of $\mathcal{B}(\mathcal{F}(\Z^d,\C^n))$ generated by
\begin{align*}
    \{a^*_{x,i}~|~x\in M,~ i\in \{1,\dots,n\}\}~.
\end{align*}
The C*-algebra $\mA \coloneq \mA_{\Z^d}$ is the CAR-algebra, which we also call the quasi-local algebra. We write $P_0(\Z^d) \coloneq \{M\subseteq \Z^d~|~|M|<\infty \}$ and call
\begin{align*}
    \mA_0 \coloneq \bigcup_{M\in P_0(\Z^d)} \mA_M \subseteq \mA
\end{align*}
the local algebra. Consequently, an operator is called quasi-local if it lies in $\mA$ and local if it lies in $\mA_0$.
For each $\varphi \in \R$ there is a unique automorphism\footnote{In the following the term {\it automorphism} 
is used in the sense of a $*$-automorphism as defined for example in~\cite{bratteliI}.}
$g_\varphi$ of $\mA$, such that
\begin{align*}
    g_\varphi(a^*_{x,i}) =  \e^{i\varphi} \, a^*_{x,i}, \quad \text{for all}~~ x\in \Z^d,~ i\in \{ 1,\dots,n \} ~.
\end{align*}
One defines the set of even quasi-local operators
\begin{align*}
    \mA^+ \coloneq \{A\in \mA~|~   g_\pi (A) = A \}
\end{align*}
and the set of gauge-invariant quasi-local operators
\begin{align*}
    \mA^N \coloneq \{A\in \mA~|~  \forall \varphi \in \R :\, g_\varphi (A) = A \} \, .
\end{align*}
Their parts in $M\subseteq\Z^d$  are denoted by $\mA^+_M \coloneq  \mA^+\cap \mA_M$ and $\mA^N_M \coloneq \mA^N\cap \mA_M$. Note that gauge-invariant operators are always even.
For disjoint regions $M_1,M_2 \subseteq\Z^d$,  $M_1\cap M_2 = \emptyset$, operators  $A\in \mA_{M_1}^+$ and $B\in \mA_{M_2}$ commute,    $[A,B] = 0$. 

Positive linear functionals of the quasi-local algebra $\w \colon \mA \to \C$ of norm $1$ are called states. In order to define quantitative notions of localization for quasi-local operators, one makes use of the fact that one can localize operators to given regions by means of the fermionic conditional expectation. 
To this end first note that  $\mA$ has a unique state $\w^{\tr}$ that satisfies
\begin{align*}
    \w^{\tr}(AB) = \w^{\tr}(BA)
\end{align*}
for all  $A,B \in \mA$, called the tracial state (e.g.\ \cite[Definition 4.1, Remark 2]{ArakiMoriya2003}).

\medskip

\begin{proposition}[{\cite[Theorem 4.7]{ArakiMoriya2003},\cite[Proposition 2.1]{wmmmt2024}}]\label{Ex+UniqueExpectation}
    For each $M\subseteq\Z^d$ there exists a unique linear map 
    \begin{align*}
        \E_M:\mA \to \mA_M\,,
    \end{align*}
    called the conditional expectation with respect to $\w^\tr$, such that
    \begin{equation}\label{eq:Conditional expectation defining property}
      \forall A\in \mA \; \;\forall B\in \mA_M \,:\quad \w^{\tr}(AB)=\w^{\tr}(\E_M(A)B) \,.
    \end{equation}
    It is unital, positive and has the properties 
    \begin{eqnarray*}
        \forall M\subseteq \Z^d\;\; \forall A,C\in \mA_M\;\; \forall B\in\mA\, : &&   \E_M \br{A\,B\,C} = A\, \E_M(B)\,C\\[1mm]
        \forall M_1,M_2 \subseteq \Z^d\,:&&   \E_{M_1} \circ \E_{M_2}  = \E_{M_1\cap M_2}\,\\[1mm]
        \forall M \subseteq \Z^d\,:&& \E_M \mA^+ \subseteq \mA^+  \quad\mbox{and}\quad \E_M \mA^N \subseteq \mA^N \\[1mm]
        \forall M \subseteq \Z^d \; \;  \forall A \subseteq \mA\,:&&  \norm{\E_M(A)}\leq \norm{A}\,.
    \end{eqnarray*}
\end{proposition}
With the help of $\E$ we can define subspaces of $\mA$ that contain operators with well-defined decay properties.

\medskip

\begin{definition}\label{def:norm}
    For $\nu \in \N_0$, $x \in \Z^d$ and $A \in \mA$ let
    \begin{align*}
        \norm{A}_{\nu,x} \coloneq \norm{A} + \sup_{k\in \N_0} \norm{A-\E_{B_k(x)}A} \, (1+k)^\nu \, ,
    \end{align*}
    where $B_k(x) \coloneq \{y\in \Z^d \, | \, \norm{x-y} \leq k \}$ is the box with side-length $2k$ around $x$ with respect to the maximum norm on $\Z^d$.
    We denote the set of all $A\in \mA$ with   $\norm{A}_{\nu,x}<\infty$ for some (and therefore all) $x\in \Z^d$ and all $\nu\in \N_0$ by $\mA_\infty$. 
\end{definition}

The following lemma quantifies the idea that, even if their supports are not strictly disjoint, operators localized in different regions have small commutators.

\medskip

\begin{lemma}\label{lem:commutator bound}
    Let $A,B \in \mA_\infty$, such that either $A$ or $B$ is even, $\nu,m \in \N_0$ and $x,y \in \Z^d$. It holds that
    \begin{align*}
        \norm{[A,B]}_{\nu,x} \leq 4^{\nu+m+3} \frac{\norm{A}_{\nu+m,y}\, \norm{B}_{\nu+m,x}}{(1+\norm{x-y})^m}\, .
    \end{align*}
    Here the norm on $\Z^d$ is again the maximum norm.
\end{lemma}
\begin{proof}
    Let $\Z^d \to \aut(\mA), \, \g \mapsto T_\g$ be the standard translation on $\mA$. We observe that 
    \begin{align*}
        \norm{[A,B]}_{\nu,x} &= \norm{T_{-x}[A,B]}_{\nu,0} = \norm{[T_{y-x}T_{-y}A,T_{-x}B]}_{\nu,0}\, .
    \end{align*}
    Now we can apply the bound from Lemma A.1 in \cite{wmmmt2024}, as all the steps of the proof work   when only one of the operators is even instead of both being gauge-invariant. We get
    \begin{align*}
        \norm{[A,B]}_{\nu,x} &\leq 4^{\nu+ m + 3} \frac{\norm{T_{-y}A}_{\nu+m,0}\, \norm{T_{-x}B}_{\nu+m,0}}{(1+ \norm{x-y})^m}\\
        &=4^{\nu+ m + 3} \frac{\norm{A}_{\nu+m,y}\, \norm{B}_{\nu+m,x}}{(1+ \norm{x-y})^m} \, . \qedhere
    \end{align*}
\end{proof}

\medskip

The relevant physical dynamics on $\mA$ is generated by densely defined derivations, which in turn are constructed from so-called interactions. 
An interaction is a map $\Phi: P_0(\Z^d) \to \mA^+$, such that $\Phi(\emptyset) = 0$ and for all $M\in P_0(\Z^d)$ it holds that $\Phi(M) \in \mA_M$, $\Phi(M)^* = \Phi(M)$, and the sum 
    \begin{equation}\label{eq:uncond}\sum_{\substack{K\in P_0(\Z^d)\\ M\cap K \neq \emptyset}} \Phi(K)
    \end{equation}
    converges unconditionally.  Note that in our definition the local terms of an interaction are always even but not necessarily gauge-invariant.

\medskip
\begin{definition}\label{norm interaction}
    For an interaction $\Phi$   and $\nu \in \N$ let
    \begin{align*}
        \|\Phi\|_{\nu} \coloneq \sup_{x\in \Z^d} \sum_{\substack{M\in
        P_0(\Z^d)\\x\in M}}(1+\mathrm{diam}(M))^\nu  \| \Phi(M)\|  ~.
    \end{align*}
    The set of interactions with finite $\norm{\cdot}_{\nu}$ for all $\nu \in \N_0$ is denoted by $\mathcal{P}_{\infty}$. 
    By $\mathrm{diam}(M)$ we mean the maximal distance of two elements in $M$ with respect to the maximum norm on $\Z^d$.\\
    Let $I\subseteq \R$ be an interval. Define $\mathcal{P}_{\infty,I}^{(k)}$ as the set of families of $\mathcal{P}_\infty$-interactions $(\Phi_s)_{s \in I}$, where for each $M \in P_0(\Z^d)$ the map $s \mapsto \Phi_s(M)$ is  $k$ times continuously differentiable and that satisfy $\sup_{s \in I}\norm{\frac{\dd^m}{\dd s^m}\Phi_s}_{\nu} < \infty$ for all $\nu \in \N_0$ and $m\leq k$.
\end{definition}

\medskip
\begin{definition}\label{def: quasi-local terms}
    For $M \in P_0(\Z^d)$ we define its center $\mathrm{C}(M) \in M$ as the point in $M$ that minimizes the distance to its center of mass $\mathrm{cm}(M)\in\R^d$. If there are multiple such points we choose the one where the standard polar-coordinate angles $$\angle(\mathrm{C}(M)-\mathrm{cm}(M))\in [0,\pi)^{d-2}\times [0,2\pi)$$ are minimal with respect to lexicographical ordering, if $d\geq 2 $ and the one larger than $\mathrm{cm}(M)$ if $d=1$.
    Let $x\in \Z^d$. We define $R_x \subseteq P_0(\Z^d)$ to be set of all finite subsets of $\Z^d$ that have their center in $x$. Given an interaction $\Phi$ we define 
    \begin{align*}
        \Phi_x \coloneq \sum_{M \in R_x} \Phi(M) \, .
    \end{align*}
\end{definition}

\medskip
\begin{remark}
    Note that for any interaction $\Phi$ and lattice point $x$ the quasi local observable $\Phi_x$ is well-defined and that $ \lVert \Phi_x \rVert_{\nu,x} \leq 3 \, \lVert \Phi \rVert_\nu $ for all $\nu \in \N_0$.
\end{remark}

\medskip 
Interactions  define derivations on $\mA$ in the following way.    
For an interaction $\Phi$ let
\begin{align*}
    \mL_{\Phi}^\circ: \mA_0 \to \mA,~ A\mapsto \sum_{M\in P_0(\Z^d)} [\Phi(M),A] \, .
\end{align*}
It follows from \cite[Proposition 3.2.22]{bratteliI} and \cite[Proposition 3.1.15]{bratteliI} that $\mL_{\Phi}^\circ$ is closable. We denote its closure by $\mL_{\Phi}$ and call it the Liouvillian of $\Phi$.

\medskip
\begin{lemma}\label{lem: sum representation of generator}
    For an interaction $\Phi \in  \mathcal{P}_\infty$, it holds that $\mA_\infty \subseteq D(\mL_{\Psi})$. Let $ A\in \mA_\infty$, then the sums
    \[\sum_{M\in P_0(\Z^d)} [\,\Phi(M) , \,A\,] \quad \text{and} \quad \sum_{x\in \Z^d} [\, \Phi_x, \, A \, ] \]
    converge absolutely and
    \[ \mL_{\Phi } \, A \;=\;  \sum_{M\in P_0(\Z^d)} [\,\Phi(M),\, A\,]  = \sum_{x\in \Z^d} [\, \Phi_x, \, A \, ] \,. \]
    For each $\nu \in \N_0$ there is a constant $c_\nu$, independent of $\Phi$ and $ A $, such that for all~$x\in \Z^d$
    \[ \norm{ \mL_{\Phi } \, A }_{\nu,x} \leq c_\nu \, \norm{\Phi}_{d+1+2\nu}  \, \norm{A}_{d+3+2\nu, x}\,.\]
\end{lemma}
\begin{proof}
    Note that the proof of Lemma 2.5 in \cite{wmmmt2024} works without change, if the local terms of the interaction are even instead of gauge-invariant and that the gauge invariance of $A$ is not used. We can therefore apply the lemma in the following way: $\mA_\infty \subseteq D(\mL_{\Psi})$ and the convergence of the first sum are directly implied. A reordering of the summation leads to the second sum. To get the estimate, denote by $T$ the family of standard translation automorphisms acting on $\mA$. By Lemma 2.5 of \cite{wmmmt2024} there is a constant $c_\nu$ independent of $x$, $\Phi$ and $A$ such that
    \begin{align*}
        \norm{ \mL_{\Phi } \, A }_{\nu,x} 
        &= \norm{  \mL_{T_{-x}\Phi } \, T_{-x}\, A }_{\nu,0}\\
        &\leq c_\nu \, \norm{T_{-x}\, \Phi}_{d+1+2\nu}  \, \norm{T_{-x}\,A}_{d+3+2\nu,0} \\
        &= c_\nu \, \norm{\Phi}_{d+1+2\nu} \, \norm{A}_{d+3+2\nu,x} \, . \qedhere
    \end{align*}
\end{proof}

\medskip

\begin{definition}\label{def:gap}
    A state $\w$ is called a \emph{ground state} of an interaction $\Phi$, if for all $A \in D(\mL_\Phi)$
    \begin{align*}
        \w(A^*\, \mL_\Phi \, A) \geq 0 \, .       
    \end{align*}
    It is called a \emph{locally-unique gapped ground state} of $\Phi$ with gap $g>0$, if for all $A \in D(\mL_\Phi)$
    \begin{align*}
        \w(A^*\, \mL_\Phi \, A) \geq g \, (\w( A^*A) - \lvert \w(A) \rvert^2)\, .
    \end{align*}
\end{definition}

\medskip

\begin{definition}
    Let $I\subseteq \R$ be an interval. A family $(\alpha_{u,v})_{(u,v) \in I^2}$ of automorphisms of $\mA$ is called a \emph{cocycle} if it satisfies
    \begin{align*}
        \forall t,u,v \in I, \quad \alpha_{t,u} \, \alpha_{u,v} =\alpha_{t,v}\, .
    \end{align*}
    We say the cocycle is generated by the family of interactions $(\Phi_{v})_{v\in I}$, if for all $A \in \mA_0$ it holds that 
    \begin{align*}
        \partial_v \, \alpha_{u,v} \, A = \alpha_{u,v} \, \i \, \mL_{\Phi_{v}} \, A  \, .   
    \end{align*}
    We call $(\alpha_{u,v})_{(u,v) \in I^2}$ locally generated if it is generated by some family of interactions in $\mathcal{P}_{\infty,I}^{(0)}$.
\end{definition}

\medskip
\begin{lemma}\label{lem: cocycles}
    Let $I \subseteq \R$ be an interval and $(\Phi_{v})_{v\in I}$ a family of interactions in $\mathcal{P}_{\infty,I}^{(0)}$. Then there exists a unique cocycle of automorphisms $(\alpha_{u,v})_{(u,v) \in I^2}$ generated by $(\Phi_{v})_{v\in I}$.
    And for all $\nu \in \N_0$ there exists an increasing function $b_\nu:\R \to \R$, that grows at most polynomially, such that
    \begin{align*}
        \norm{\alpha_{u,v} \, A}_{\nu,x} \leq b_\nu(|v-u|) \norm{A}_{\nu,x} \,   \quad \text{ for all } x\in \Z^d, \, A\in \mA_\infty,\, u,v\in I \,.
    \end{align*}
    Furthermore, for all  constants $C>0$ the function $b_\nu$ can be chosen uniformly for all $(\Phi_v)_{v\in I}$ \linebreak with $\sup_{s\in I} \lVert \Phi_v \rVert_{4\nu+9d+4} < C$.
\end{lemma}
The proof of this last lemma, which relies on the Lieb-Robinson bounds from \cite{TeufelWessel25}, is given in Appendix \ref{sec: Lieb-Robinson bounds}.

\section{Main Results} \label{sec:results}

Let $I \subseteq \R $ be an interval, $(H_s)_{s\in I}$ a family of interactions and $(\w_s)_{s\in I}$ a family of states. We make the following assumptions:

\begin{enumerate}
    \item[(i)] \label{Assumption: Hs is in B infty} The family $(H_s)_{s \in I}$ lies in $\mathcal{P}_{\infty , I}^{(1)} $.
    \item[(ii)] There exists a $g>0$, such that for each $s\in I$, the state  $\w_s$ is a locally-unique gapped ground state of $H_s$ with gap $g$.
    \item[(iii)] \label{Assumption: Cont. diffble} For all $A\in \mA_\infty$ it holds that $s \mapsto \w_s(A)$ is differentiable and there is a $\nu \in \N_0$, such that for some (and therefore all) $x \in  \Z^d$, there exists a constant $C_{\nu,x}$ with
    \begin{align*}
        |\Dot{\w}_s(A)|\leq C_{\nu,x} \, \lVert A\rVert_{\nu,x} \quad \forall A \in \mA_\infty, \, s\in I \, .
    \end{align*}
\end{enumerate}
Note that our conditions on the interaction $H_s$ are less restrictive compared to those in  \cite{moon2019automorphic}.
\\
Our third assumption is equivalent to the condition that $\Dot{\w}_s$ is a continuous functional on the Fr\'echet space $\mA_\infty$. This assumption is stronger than the corresponding one in \cite{moon2019automorphic}, where continuity is only required with respect to a norm $\|\cdot\|_{f,x}$ defined as in Definition~\ref{def:norm}, but with $(1+k)^{-\nu}$ replaced by $f(k)$ for a fixed  decreasing stretched exponential $f$.

\medskip
\begin{definition}\label{def: differentiable path of gapped systems}
    Let $I \subseteq \R $ be an interval, $(H_s)_{s\in I}$ a family of interactions and $(\w_s)_{s\in I}$ a family of states such that the above assumptions are satisfied. 
    Then we call the pair $((\w_s)_{s\in I}, (H_s)_{s\in I})$ a \emph{differentiable path of gapped systems}.
\end{definition}

We can now formulate our main theorem. It states that  for any differentiable path of gapped systems there is a locally generated cocycle of automorphisms, called the spectral flow or the parallel transport, that transports the ground states along the path.

\medskip

\begin{theorem}\label{Theorem: main theorem}
    Let $((\w_s)_{s\in I}, (H_s)_{s\in I})$ be a differentiable path of gapped systems. Then the cocycle of automorphisms $(\alpha_{u,v})_{(u,v) \in I^2}$ generated by the family of interactions $(-\mI_s(\Dot{H}_s))_{s\in I}$ (defined in Appendix \ref{sec: off-diag})  satisfies
    \begin{align*}
        \w_t = \w_s \circ \alpha_{s,t}\qquad\mbox{for all $s,t\in I$.}
    \end{align*}
\end{theorem}
\begin{proof}[Proof of Theorem \ref{Theorem: main theorem}]
    Let $A \in \mA_\infty$. By Lemma \ref{lem: od-splitting} we can split $A$ into its diagonal and off-diagonal parts with respect to $\w_s$ (see Appendix \ref{sec: off-diag} for precise definitions), and hence
    \begin{align*}
        \Dot{\w}_s(A) 
        & = \Dot{\w}_s(A\D[s])+ \Dot{\w}_s(A\OD[s])\, .
    \end{align*}
     Proposition \ref{Proposition: vanishing derivative} states that the diagonal part does not contribute, i.e.\ that $\Dot{\w}_s(A\D[s])=0$. Thus, by inserting the definition of $A\OD[s]$, we obtain
    \begin{align*}
        \Dot{\w}_s(A)
        \,=\, \Dot{\w}_s(A\OD[s])
         \,=\, \Dot{\w}_s(-\i \, \mL_{H_s} \, \mI_s(A))\, .
    \end{align*}
    Next we employ Lemma \ref{lem: derivative switch}, which uses the fact that the ground state $\w_s$ is stationary with respect to $H_s$, to move the derivative from the state to the interaction,  
    \begin{align*}
        \Dot{\w}_s(A)
        & = \w_s(\i\,\mL_{\Dot{H}_s} \, \mI_s \,A )\\
        & = \int_\R \dd t \, W(t)\, \w(\i \, \mL_{\Dot{H}_s} \, \e^{\i t \mL_{H_s}} \,A)\, ,
    \end{align*}
    where we also inserted the definition of the inverse Liouvillian $\mI_s$. Now we use the stationarity of the ground state again to insert the inverse time evolution into the expression and apply Lemma~\ref{lem: Inverse Liouvillian on derivations} to arrive at
    \begin{align*}
        \Dot{\w}_s(A)
        & = \int_\R \dd t \, W(t)\, \w(\i \, e^{- \i t \mL_{H_s}}\, \mL_{\Dot{H}_s} \, \e^{\i t \mL_{H_s}} \,A)\\
        & = - \w_s( \i\, \mL_{\mI_s(\Dot{H}_s)} \, A)\, .
    \end{align*}   
    Consider now the expression $\w_s(\alpha_{s,t} \, A )$. For $s=t$ this equals  $\w_t(A)$. Taking the derivative with respect to $s$ yields
    \begin{align*}
        \partial_s\,\w_s(\alpha_{s,t} \, A ) 
        & = \Dot{\w}_s(\alpha_{s,t} \, A ) + \w_s(\partial_s\,\alpha_{s,t} \, A ) \\
        & = -\Dot{\w}_s(\i\, \mL_{\mI_s(\Dot{H}_s)}\,\alpha_{s,t} \, A ) + \w_s(\i\, \mL_{\mI_s(\Dot{H}_s)}\, \alpha_{s,t} \, A )= 0 \, .
    \end{align*}
    Therefore we have $\w_s(\alpha_{s,t} \, A )= \w_t(A)$ for all $A \in \mA_\infty$. By the density of $\mA_\infty$ in $\mA$ it follows that $\w_s \circ \alpha_{s,t} = \w_t$.
\end{proof}

The next theorem is a version of the so-called Goldstone theorem. For self-adjoint Hamiltonians acting on Hilbert spaces it is a simple observation that any unitary that commutes with $H$ leaves invariant the spectral subspaces of $H$ and thus, in particular, maps any non-degenerate eigenspace into itself. For locally-unique gapped ground states of interactions on the quasi-local algebra, a similar statement holds for automorphisms that are generated by quasi-local interactions. Proving this is less obvious and for spin systems with  finite-range interactions this was   established in \cite{wreszinski1987charges}. Here we obtain Goldstone's theorem for $\mathcal{P}_\infty$ interactions and also for lattice fermion systems as a simple consequence of the previous theorem.

\medskip
\begin{theorem}\label{thm: invariance of locally unique gs}
    Let $H$ be a $\mathcal{P}_\infty$-interaction and $\w_0$ a locally-unique gapped ground state of $H$. Further, let $I\subseteq \R$ be an interval and $(\beta_{u,v})_{(u,v) \in I^2}$ a cocycle of automorphisms generated by $(\Phi_v)_{v\in I} \in \mathcal{P}_{\infty,I}^{(0)}$, such that for all $ u,v \in I$ and $A\in \mA_0$ it holds that  $\beta_{u,v} \, \mL_H\, A= \mL_H\,\beta_{u,v} \, A$. 
    
    Then $\w_0 \circ \beta_{u,v} = \w_0$ for all $u,v \in I$.
\end{theorem}
\begin{proof}
    Let $u \in I$. We will show that the family of states $(\w_0\circ \beta_{u,v})_{v\in I}$ together with the constant family of interactions $(H_v\equiv H)_{v\in I}$ form a differentiable path of gapped systems, i.e.\  that the assumptions of Theorem~\ref{Theorem: main theorem} are satisfied. Then Theorem~\ref{Theorem: main theorem} implies that the cocycle $\alpha_{v,w}$ generated by $(-\mI_s(\Dot{H}))_{s\in I}$ satisfies
    \[
    \w_0\circ \beta_{u,v} = \w_0\circ \beta_{u,w} \circ \alpha_{w,v}\,.
    \]
    But $\Dot{H}=0$, which implies  $\alpha_{w,v}=\mathrm{id}$ for all $w,v\in I$ and thus $\w_0\circ \beta_{u,v} = \w_0\circ \beta_{u,u} = \w_0$ for all $v\in I$.

    It remains to check the assumptions of Theorem~\ref{Theorem: main theorem} for the above path of gapped systems. 
    Since $\w_0$ is a locally-unique gapped ground state there is a $g>0$ such that 
    \begin{align*}
        \forall A \in D(\mL_{H}), \quad \w_0(A^* \, \mL_H \, A) \geq g \, (\w_0(A^*\, A) - \lvert \w_0(A) \rvert^2) \, .
    \end{align*}
    Let $A \in \mA_0$ and $v \in I$. Applying the above condition to $\alpha_{u,v} \, A \in D(\mL_H)$ yields
    \begin{align*}
        \w_0((\alpha_{u,v}\,A)^* \, \mL_H \, \alpha_{u,v}\,A) \geq g \, (\w_0((\alpha_{u,v}\,A)^*\, \alpha_{u,v}\,A) - \lvert \w_0(\alpha_{u,v}\,A) \rvert^2)\, .
    \end{align*}
    We use that by assumption the derivation $\mL_H$ commutes with $\alpha_{u,v}$ to get 
    \begin{align*}
        (\w_0\circ \alpha_{u,v})(A^* \, \mL_H \, A) \geq g \, (\w_0\circ \alpha_{u,v})(A^*\, A) - \lvert (\w_0 \circ \alpha_{u, v})(A) \rvert^2\, .
    \end{align*}
    Since $\mA_0$ is a core of $\mL_H$ this implies that $\w_0 \circ \alpha_{u,v}$ is a locally unique gapped ground state of $H$ with gap $g$ for all $v \in I$. Looking at the derivative with respect to $v$, we find by Lemma \ref{lem: sum representation of generator}, that there exists a $c>0$ such that
    \begin{align*}
        \forall A\in \mA_\infty, \quad \lvert \partial_v \, \w_0( \alpha_{u,v}\, A) \rvert  = \lvert \w_0( \alpha_{u,v} \, \i \, \mL_{\Phi_v} \, A) \rvert \leq c\, \lVert \Phi_v \rVert_{d+1} \, \lVert A \rVert_{d+3,0}\, ,
    \end{align*}
    where $(\Phi_v)_{v\in I} \in \mathcal{P}_{\infty,I}^{(0)}$ generates the cocycle. Thus, $((\w_0\circ \alpha_{u,v})_{v \in I}, (H)_{v\in I})$ is a differentiable path of gapped systems and the assumptions of Theorem \ref{Theorem: main theorem} are satisfied. 
\end{proof}
Note that any automorphism which leaves a state $\w$ invariant is unitarily implementable in its GNS representation. Therefore, the previous theorem also implies that continuous symmetries can be implemented unitarily in the GNS representation of a locally-unique gapped ground state.

As a simple but important example, we find that locally-unique gapped ground states of gauge-invariant interactions are gauge-invariant.

\medskip
\begin{corollary}\label{cor: gauge-invariance}
    Let $H \in \mathcal{P}_\infty$ be a gauge-invariant interaction (i.e. all local terms lie in $\mA_0^N$) and let $\w_0$ be a locally-unique gapped ground state of $H$. Then $\w_0$ is a gauge-invariant state:
    \begin{align*}
        \forall \varphi \in \R, \quad \w_0 \circ g_\varphi = \w_0 \, .
    \end{align*}
\end{corollary}
\begin{proof}
    Since $H$ is gauge invariant, the family of gauge automorphisms $(g_\varphi)_{\varphi \in \R}$, which is generated by the number  operator $N\in \mathcal{P}_\infty$, defined as
    \begin{align*}
        N(M) \coloneq \begin{cases}
            n_x\, , \quad &M= \{x\},\,  x\in \Z^d\\
            0\, , \quad &\lvert M \rvert \neq 1
        \end{cases}
    \end{align*}
    for all $M \in P_0(\Z^d)$, satisfies the assumptions of Theorem \ref{thm: invariance of locally unique gs}.
\end{proof}

\section{The Parallel Transport Condition}
\label{sec:partrans}

In this section, we prove the most essential ingredient used in the proof of Theorem \ref{Theorem: main theorem}, namely the parallel transport condition. It tells us that the derivative of the ground state applied to the diagonal part of an operator vanishes. We first give the structure of the argument, which is a streamlined version of the strategy developed in   \cite{moon2019automorphic}, and then provide technical lemmas to complete the proof. 

\medskip
\begin{proposition}\label{Proposition: vanishing derivative}
    Let $((\w_s)_{s\in I}, (H_s)_{s\in I})$ be a differentiable path of gapped systems and $A\in \mA_\infty$. Then the parallel transport condition holds:
    \begin{align*}
        \forall s \in I, \quad \Dot{\w}_s(A\D[s])=0  \, .
    \end{align*}
\end{proposition}
\begin{proof}
    We start by using \cite[Lemma 3.2]{moon2019automorphic}. This tells us that for all $s \in I$,  $A,B \in \mA$ it holds that
    \begin{align*}
        \w_s(B^*\, A\D[s]) = \w_s(B^*) \, \w_s(A)\, . \label{eq: factorization} \numberthis
    \end{align*}
    From this we conclude that $ (A\D[s] - \w_s(A))^* $ lies in the left vanishing ideal of $\w_s$ 
    \[ \Lker(\w_s) \coloneq \{ C \in \mA \, |\, \w_s(C^*\,C) =0 \}\, .\]
    Let $s_0 \in I$ and let $\nu \in \N_0$ be such that $\Dot{\w}_{s_0}$ is continuous with respect to $\lVert \, \cdot \, \rVert_{\nu,0}$. Applying Lemma~\ref{Lemma: sequence} to $(A\D[s_0]-\w_{s_0}(A))^*$ and $\nu$ gives us an approximating sequence $(u_N)_{N\in\N}$ such that
    \begin{align*}
        \lVert(1-u_N)^* (A\D[s_0] - \w_{s_0}(A)) \rVert_{\nu,0}  \xrightarrow{N\to \infty} 0, \numberthis \label{eq: uN convergence}
    \end{align*}
    and 
    \begin{align*}
        \mathrm{dist}(u_N,\Lker(\w_{s_0})) \coloneq \inf_{C \in \Lker(\w_{s_0})} \lVert u_N -C \rVert \xrightarrow{N\to \infty} 0.  \numberthis \label{eq: distance vanishes}
    \end{align*}
    By Equation \eqref{eq: factorization} it holds that
    \begin{align*}
       \forall s\in I, \quad \w_s(u_N^*\, (A\D[s] - \w_s(A)) \,) = 0 \, .
    \end{align*}
    Taking a derivative with respect to $s$ yields
    \begin{align*}
        0 &= \partial_s\, \w_s(u_N^*\, (A\D[s] - \w_s(A)) \,) \\
        & = \Dot{\w}_s(u_N^*\, (A\D[s] - \w_s(A)) \,) + \w_s(u_N^*\, \partial_s\, (A\D[s] - \w_s(A) \, ) \,)\, ,
    \end{align*}
    where we used that by Lemma \ref{Lemma: Differential off diagonal} the function $s \mapsto A\D[s] - \w_s(A)$ is differentiable with respect to $\lVert \, \cdot \, \rVert$. Setting $s=s_0$ leaves us with
    \begin{align*}
        0 = \Dot{\w}_{s_0}(u_N^*\, (A\D[s_0] - \w_{s_0}(A)) \,) + \w_{s_0}(u_N^*\, (\partial_s\, A\D[s] - \Dot{\w}_{s_0}(A))|_{s=s_0})   \, .
    \end{align*}
    Due to the continuity of $\Dot{\w}_{s_0}$ and \eqref{eq: uN convergence}, the first term converges to $\Dot{\w}_{s_0}(A\D[s_0]-\w_{s_0}(A))=\Dot{\w}_{s_0}(A\D[s_0])$ as $N \to \infty$. The second term vanishes in this limit, since for any $B\in \mA$ we have
    \begin{align*}
        \lvert \w_{s_0}(u_N^*\, B ) \rvert
        & = \inf_{C \in \Lker(w_{s_0})} \lvert \w_{s_0}((u_N - C  + C)^* \, B ) \rvert \\
        & \leq \inf_{C \in \Lker(w_{s_0})}  \lVert u_N - C \rVert \, \lVert B  \rVert +  \lvert \w_{s_0}(C^* \, B ) \rvert \, \\
        & = \inf_{C \in \Lker(w_{s_0})}  \lVert u_N - C \rVert \, \lVert B  \rVert \, ,
    \end{align*}
    which vanishes for $N \to \infty$, due to \eqref{eq: distance vanishes}. Thus, we are left with the statement 
    \begin{equation*}
        \Dot{\w}_{s_0}(A\D[s_0]) = 0 \, . \qedhere
    \end{equation*}
\end{proof}

In the next lemma, we will construct a sequence $u_N$, whose distance from $\Lker(\w_s)$ converges to zero and when multiplied by an element of $\Lker(\w_s)$ it converges to this element in an appropriate sense. A corresponding lemma was shown in \cite[Lemma 3.6]{moon2019automorphic}.

\medskip
\begin{lemma}\label{Lemma: sequence}
    Let $((\w_s)_{s\in I}, (H_s)_{s\in I})$ be a differentiable path of gapped systems, let $\nu \in \N_0$, $s\in I$ and $A\in \mA_\infty \cap \Lker(\w_s)$, where $\Lker(\w_s) \coloneq \{ C \in \mA \, |\, \w_s(C^* \, C)=0 \}$ is the left vanishing ideal of $\w_s$.
    There exists a sequence $(u_N)_{N\in \N}$ in $\mA$ such that 
    \begin{align*}
        \mathrm{dist}(u_N,\Lker(\w_{s})) \coloneq \inf_{C \in \Lker(\w_{s})} \lVert u_N -C \rVert \xrightarrow{N\to \infty} 0. 
    \end{align*}
    and 
    \begin{align*}
        \lVert A \, (1- u_N) \rVert_{\nu,0}  \xrightarrow{N\to \infty} 0,
    \end{align*}
\end{lemma}
\begin{proof}
    Let $A \in \Lker( \w_s ) \cap \mA_\infty$ and set
    \begin{align*}
        h(N) &\coloneq (1 + N)^{-( 2 \nu + 1 )}\\
        u_N  &\coloneq ( h(N) + \E_{B_N(0)} (A^*A))^{-1} \, \E_{B_N(0)}(A^*A) \, .
    \end{align*}
    Note that elements of the form $c+B$, where $c>0$ and $B\in \mA$ is positive, are always invertible in $\mA$ with $\lVert(c+B)^{-1}\rVert \leq \frac{1}{c}$. To prove the first statement we consider
    \begin{align*}
        \hspace{2em}&\hspace{-2em} \lVert u_N  - (h(N) + A^*A)^{-1} \, A^*A \rVert \\
        & = \lVert u_N  -  A^*A \, (h(N) + A^*A)^{-1}  \rVert \\
        & = \lVert ( h(N) + \E_{B_N(0)} (A^*A))^{-1} \E_{B_N(0)}(A^*A) - A^*A  \, (h(N) + A^*A)^{-1} \rVert\, .
    \end{align*}
    Multiplicatively  inserting a one into both terms and using that $\E_{B_N(0)} (A^*A)$ commutes with $A^* A$ yields
    \begin{align*}
        \hspace{2em}&\hspace{-2em} \lVert u_N  - A^*A  \, (h(N) + A^*A)^{-1} \rVert \\
        & = \lVert (h(N) + \E_{B_N(0)} (A^*A))^{-1} h(N) \, (A^*A-\E_{B_N(0)}(A^*A))  \, ( h(N) + A^*A)^{-1} \rVert \\
        & \leq h(N)^{-1} \lVert  A^*A-\E_{B_N(0)}(A^*A) \rVert  \\
        &\leq \frac{h(N)^{-1}}{(1+N)^{2\nu + 2}} \lVert A^*A \rVert_{2\nu + 2, 0} \xrightarrow{N \to \infty} 0 \, .
    \end{align*}
    Since $\Lker( \w_s )$ is a left ideal, we have $(h(N) + A^*A)^{-1} A^*A \in \Lker( \w_s )$ and hence the first statement immediately follows.

    For the second claim, we first observe that 
    \begin{align*}
        \lVert A (1 - u_N) \rVert^2 
        & \leq \lVert (1 - u_N) \, (A^*A - \E_{ B_N(0)}(A^*A)) \, (1 - u_N) \rVert \\
        &\quad + \lVert (1 - u_N) \, \E_{ B_N(0)}(A^*A) \, (1 - u_N) \rVert\, .
    \end{align*}
    Using that 
    \begin{align*}
        \lVert 1-u_N \rVert &= \lVert (h(N) + \E_{B_N(0)}(A^*A))^{-1} \, h(N) \rVert \leq 1
    \end{align*}
    and
    \begin{align*}
        (1-u_N)\, \E_{B_N(0)}(A^*A) = h(N)\, u_N\, 
    \end{align*}
    we find 
    \begin{align*}\label{eq: u_N square norm}
        \lVert A (1 - u_N) \rVert^2
        &\leq h(N) \lVert A^*A \rVert_{2\nu + 1 , 0 } + h(N) \, . \numberthis      
    \end{align*}
    In the next step, we consider two cases. First, for $k>N$, we have
    \begin{align*}
        &\lVert A (1 - u_N ) - \E_{B_k(0)} \, ( A ( 1 - u_N ) ) \rVert \, ( 1 + k)^\nu 
        \leq \lVert A -\E_{B_k(0)} \, A \rVert \, (1 + k)^\nu 
        \\
        & \qquad \leq \lVert A \rVert_{ \nu + 1 , 0}\frac{1}{1+k} \xrightarrow{N \to \infty} 0 \, .
    \end{align*}
    And for $k\leq N$, with \eqref{eq: u_N square norm} we get
    \begin{align*}
        \hspace{2em}&\hspace{-2em} \lVert A (1 - u_N) - \E_{B_k(0)} ( A ( 1 - u_N ) ) \rVert \, ( 1 + k )^\nu
        \\
        & \leq 2 \lVert A ( 1 - u_N )\rVert \, ( 1 + N )^\nu
        \\
        & \leq 2 \left( h(N) \,  \lVert  A \rVert_{2 \nu + 1 , 0} + h(N) \right)^{1 / 2} (1 + N)^\nu \xrightarrow{N \to \infty} 0 \, .
    \end{align*}
    Combining the two cases gives the second statement.
\end{proof}

In the next lemma, we show that $A\D[s]$ is differentiable and also give its derivative explicitly.

\medskip
\begin{lemma}\label{Lemma: Differential off diagonal}
    Let $((\w_s)_{s\in I}, (H_s)_{s\in I})$ be a differentiable path of gapped systems and $A\in \mA_\infty$. Then the map $I\ni s \mapsto A\D[s]$ is differentiable (with respect to $\lVert \, \cdot \, \rVert$) and
    \begin{align*}
        \frac{\d}{\d s}A\D[s]=V_s(A) \, ,
    \end{align*}
    where $V_s$ is given by
    \begin{align*}
        V_s(A) \coloneq \i \int_\R \d t \, w(t) \int_0^t\d u \, \e^{\i(t-u) \mL_{H_s} } \, \mL_{ \Dot{H}_s } \, \e^{ \i  u \mL_{H_s} } \, A \, .
    \end{align*}
\end{lemma}
\begin{proof}
    By the fundamental theorem, we have
    \begin{align*}
        ( \e^{ \i t \mL_{ H_{s+h} }} - \e ^ {\i t\mL_{H_s}}) \, A = \i \int_0^t \d u \, \e^{\i  ( t - u ) \mL_{H_{s+h}}} \, ( \mL_{H_{s+h}} - \mL_{H_s} ) \, \e^{ \i  u \mL_{H_s}}  \, A \, .
    \end{align*}
    Multiplying by $w(t)$ and integrating over $\R$ yields
    \begin{align*}
        A\D[(s+h)] - A\D[s] = \i \int_{ \R } \d t \, w(t) \int_0 ^ t \d u \, \e^{ \i ( t - u ) \mL_{ H^{ s + h }}} \, ( \mL_{H_{ s + h }} - \mL_{ H_s } ) \, \e^{ \i u \mL_{H_s}} \, A \, .
    \end{align*}
    Now the statement follows by calculating
    \begin{align*}
        \lVert \frac{ 1 }{ h } &( A\D[ ( s + h ) ] - A\D[ s ]) - V_s(A) \rVert \leq
        \\
        \leq &\; \i \int_{\R} \d t \, w(t) \int_0^t \d u \, \left( \left\| \e^{ \i ( t - u ) \mL_{ H_s }} \frac{ \mL_{ H_{s + h} } - \mL_{ H_s }}{h} e^{ \i u \mL_{H_s}} \, A - \e^{ \i ( t - u ) \mL_{ H_s } } \mL_{ \Dot{H}_s } \, \e^{ \i u \mL_{ H_s }} \, A \right\| \right)
        \\
        \leq &\;\i \int_{\R} \d t \, w( t ) \int_0^t \d u \,  \left\| \left( \frac{ \mL_{ H_{s + h} } - \mL_{ H_s }}{h} - \mL_{ \Dot{H}_s } \right) \, \e^{ \i u\mL_{ H_s }} \, A \right\|
        \\
        &\qquad + \i \int_{\R} \d t \, w( t ) \int_0^t \d u \, \left\| \left( \e^{ \i ( t - u ) \mL_{ H_{s + h} }} - \e^{ \i ( t -u ) \mL_{ H_s }}\right) \, \mL_{ \Dot{H}_s } \e^{ \i u \mL_{ H_s }} \, A \right\|  \, .
    \end{align*}
    The first integrand vanishes in the limit $h \to 0$ by Lemma \ref{lem: derivative of Liouvillian}. Since, by Lemmas \ref{lem: sum representation of generator} and \ref{lem: cocycles}, we have the bound
    \begin{align*}
        \left\| \left( \frac{ \mL_{ H_{s + h} } - \mL_{ H_s }}{h} - \mL_{ \Dot{H}_s } \right) \, \e^{ \i u\mL_{ H_s }} \, A \right\| 
        & \leq b(u) \sup_{v\in I} \lVert \Dot{H}_v \rVert_{d+5} \, \lVert A \rVert_{d+3}
        \, ,
    \end{align*}
    with an increasing at most polynomially growing function $b$, the first integral vanishes in the limit $h \to 0$ by dominated convergence.

    For the second term, we have, by Duhamel's formula, that for any $B \in \mA_\infty$
    \begin{align*}
        \lVert ( \e^{ \i t \mL_{ H_{s + h} }} - \e^{ \i t \mL_{ H_s }}) \, B \rVert &\leq \int_0^h \d u \, \lVert \partial_u \, \e^{\i t \mL_{H_{s+u}}} \, B \rVert
        \\
        &\leq  \int_0^h \d u \, \int_0^t \d v \, \lVert \i  \, \e^{\i  v \mL_{H_{s+u}}} \, \mL_{\Dot{H}_{s+u}} \, \e^{\i (t  - v) \mL_{H_{s+u}}} \, B \rVert \, ,
    \end{align*}
    which vanishes in the limit $h \to 0$ by the Lemmas \ref{lem: sum representation of generator} and \ref{lem: cocycles}.
\end{proof}

\appendix

\section{Lieb--Robinson Bounds and Proof of Lemma \ref{lem: cocycles}}\label{sec: Lieb-Robinson bounds}

\begin{proof}[Proof of Lemma \ref{lem: cocycles}]
    Corollary 5.2 from \cite{Bru2017} shows that a family of interactions $(\Phi_v)_{v\in I} \in \mathcal{P}_{\infty,I}^{(0)}$ always generates a unique cocycle of automorphisms. In the remaining proof, we follow the standard strategy of showing existence of infinite-volume dynamics from Lieb--Robinson bounds (see for example \cite{Bru2017}) and additionally show that the rate of convergence is super-polynomial with our assumptions. For this we will employ the improved Lieb-Robinson bounds for fermions of \cite{TeufelWessel25} (see also \cite{EMN2020} for similar bounds for spin systems), which provide us with an algebraic light cone.
    
    Let $k \in \N_0$ and $z\in \Z^d$. We first define a finite-volume approximation of $\Phi_v$ on $B_k(z)$ as the interaction $\Phi_{v,k}$, defined by
    \begin{align*}
        \Phi_{v,k}(M) \coloneq \E_{B_{k/2}(x)}\, \Phi_v(M)   \, 
    \end{align*}
    for all $M$ with center in $x \in B_{k/2}(z)$  and $\Phi_{v,k}(M) = 0 $ otherwise. Let $(\alpha^k_{u,v})_{(u,v)\in I^2} $ be the family of automorphisms generated by $(\Phi_{v,k})_{v\in I}$. Note that these automorphisms map $\mA_{B_k(z)}$ to itself, since all local terms of $\Phi_{v,k}$ lie in $\mA_{B_k(z)}$. 
    
    Let $A \in \mA_\infty$ and $\nu \in \N_0$. Our goal is to show a bound that decays faster than $\frac{1}{(1+k)^\nu}$ in $k$ while growing at most polynomially in $\lvert v - u \rvert $ for the quantity
    \begin{align*}
        \lVert (1-\E_{B_k(z)}) \, \alpha_{u,v} \, A \rVert 
        \leq 2 \, \lVert (1 - \E_{B_{k/4}(z)}) \, A \rVert + \lVert (1-\E_{B_k(z)}) \, \alpha_{u,v} \, \E_{B_{k/4}(z)}\, A \rVert   \, .
    \end{align*}
    The first term can be bounded by $ \frac{2\, \lVert A \rVert_{\nu,z}}{(1+k/4)^\nu}$. We focus on the second term:
    \begin{align*}
        \lVert (1-\E_{B_k(z)}) \, \alpha_{u,v} \, \E_{B_{{k/4}}(z)}\, A \rVert 
        & = \lVert (1-\E_{B_k(z)}) \, (\alpha_{u,v} -\alpha^k_{u,v}) \, \E_{B_{{k/4}}(z)}\, A \rVert  \\
        & \leq  2 \, \lVert (\alpha_{u,v} -\alpha^k_{u,v}) \, \E_{B_{{k/4}}(z)}\, A \rVert \\
        & \leq 2 \int_{u}^v \dd s \, \lVert \partial_s \, \alpha_{u,s} \, \alpha^k_{s,v} \, \E_{B_{{k/4}}(z)}\, A \rVert \\
        & = 2 \int_{u}^v \dd s \, \lVert \alpha_{u,s} \, (\mL_{\Phi_s} - \mL_{\Phi_{s,k}}) \, \alpha^k_{s,v} \, \E_{B_{{k/4}}(z)}\, A \rVert\, .
    \end{align*}
    Note that $\alpha^k_{s,v} \, \E_{B_{{k/4}}(z)}\, A$ lies in the domain of $\mL_{\Phi_s}$, since it is a local operator. We continue our calculation by using that the first automorphism does not change the norm and inserting the sum representations for both derivations (Lemma \ref{lem: sum representation of generator}), resulting in
    \begin{align*}
        \lVert (1-\E_{B_k(z)}) \, \alpha_{u,s} \, \E_{B_{{k/4}}(z)}\, A \rVert
        & \leq 2 \int_{u}^v \dd s \, \lVert (\mL_{\Phi_s} - \mL_{\Phi_{s,k}}) \, \alpha^k_{s,v} \, \E_{B_{{k/4}}(z)}\, A \rVert\\
        & \leq 2 \int_{u}^v \dd s \sum_{x \in B_{k/2}(z)} \lVert [( 1- \E_{B_{k/2}(x)}) \, \Phi_{s,x}, \,  \alpha^k_{s,v} \, \E_{B_{k/4}(z)}\, A] \rVert\\
        & \quad + 2 \int_{u}^v \dd s \sum_{x \in \Z^d  \setminus B_{k/2}(z)} \lVert [\Phi_{s,x}, \,  \alpha^k_{s,v} \, \E_{B_{k/4}(z)}\, A] \rVert\, .
    \end{align*}
    The first of these terms is easily handled:
    \begin{align*}
        2 \int_{u}^v \dd s \sum_{x \in B_{k/2}(z)} \lVert [( 1- \E_{B_{k/2}(x)}) \, \Phi_{s,x}, \,  \alpha^k_{s,v} \, \E_{B_{k/4}(z)}\, A] \rVert
        & \leq 4 \, \lvert v - u \rvert \sum_{x\in B_{k/2}(z) }  \frac{K}{(1+k/2)^{\nu+d}} \, \lVert A \rVert \, ,
    \end{align*}
    where we defined $K \coloneq \sup_{s\in I} \lVert \Phi_s \rVert_{4\nu+9d+4} $.
    We further focus on the second term:
    \begin{align*}
        2 \int_{u}^v \dd s &\sum_{x \in \Z^d  \setminus B_{k/2}(z)} \lVert [\Phi_{s,x}, \,  \alpha^k_{s,v} \, \E_{B_{{k/4}}(z)}\, A] \rVert \\
        &= 2 \int_{u}^v \dd s \sum_{x \in \Z^d  \setminus B_{k/2}(z)} \lVert [\, \alpha^k_{v,s} \, \Phi_{s,x}, \,  \E_{B_{{k/4}}(z)}\, A\, ] \rVert\\
        & \leq 2 \int_{u}^v \dd s \sum_{x \in \Z^d  \setminus B_{k/2}(z)} \lVert [\, \alpha^k_{v,s}  \, (1-\E_{B_{\lVert x-z \rVert/4}(x)}) \, \Phi_{s,x}, \,  \E_{B_{{k/4}}(z)}\, A\, ] \rVert\\
        & \quad + 2 \int_{u}^v \dd s \sum_{x \in \Z^d  \setminus B_{k/2}(z)} \lVert [\, \alpha^k_{v,s}  \, \E_{B_{\lVert x-z \rVert/4}(x)} \, \Phi_{s,x}, \,  \E_{B_{{k/4}}(z)}\, A\, ] \rVert
    \end{align*}
    Again the first term is easily handled:
    \begin{align*}
        \hspace{2em}&\hspace{-2em}  2 \int_{u}^v \dd s \sum_{x \in \Z^d  \setminus B_{k/2}(z)} \lVert [\, \alpha^k_{v,s}  \, (1-\E_{B_{\lVert x-z \rVert/4}(x)}) \, \Phi_{s,x}, \,  \E_{B_{{k/4}}(z)}\, A\, ] \rVert\\
        & \leq 4 \, \lvert v - u \rvert \sum_{x \in \Z^d \setminus B_{k/2}(z)} \frac{K}{(1+k/8)^\nu \, (1+ \lVert x -z \rVert/4 )^{d+1}}   \lVert A \rVert \, .
    \end{align*}
    This leaves us with the final expression, on which we use the Lieb--Robinson bounds from Proposition \ref{prop: LiebRobinson} (choosing the exponent of the decay to be $2\nu + 4d +2$). This gives us an increasing and at most polynomially growing function $f$ and a constant $c>0$ such that
    \begin{align*}
        \hspace{-4em}&\hspace{4em} 2 \int_{u}^v \dd s \sum_{x \in \Z^d  \setminus B_{k/2}(z)} \lVert [\, \alpha^k_{v,s}  \, \E_{B_{\lVert x-z \rVert/4}(x)} \, \Phi_{s,x}, \,  \E_{B_{{k/4}}(z)}\, A\, ] \rVert
        \\
        & \leq   \sum_{x \in \Z^d  \setminus B_{k/2}(z)}  \frac{2 \, K \,\lvert v - u \rvert\,  \lVert A \rVert  \, \lvert B_{k/4}(z) \rvert \, f(\lvert v-u \rvert) }{(1+\max(0, (3\lVert x - z \rVert/4-k/4)^{\frac{1}{2}} - c \,\lvert v - u \rvert))^{2\nu+4d+2}}\, , 
        \end{align*}
        where we used that the distance between the two supports is $3 \lVert x -z \rVert /4-k/4$. The distance can be estimated in two different ways, yielding
        \begin{align*}
        \hspace{-4em}& 2 \int_{u}^v \dd s \sum_{x \in \Z^d  \setminus B_{k/2}(z)} \lVert [\, \alpha^k_{v,s}  \, \E_{B_{\lVert x-z \rVert/4}(x)} \, \Phi_{s,x}, \,  \E_{B_{{k/4}}(z)}\, A\, ] \rVert
        \\
        & \leq    \sum_{x \in \Z^d  \setminus B_{k/2}(z)}  \frac{2 \,K \,\lvert v - u \rvert \, \lVert A \rVert  \, \lvert B_{k/4}(z) \rvert \, f(\lvert v-u \rvert)}{(1+\max(0, (k/8)^\frac{1}{2} - c \lvert v - u \rvert))^{2\nu+2d}}  \frac{1}{(1+\max(0, (\lVert x -z \rVert/4)^{\frac{1}{2}} - c \lvert v - u \rvert))^{2d+2}} \, .
    \end{align*}
    Using Lemma \ref{lem: function estimate}, we now find 
    \begin{align*}
          \sum_{x \in \Z^d  \setminus B_{k/2}(z)}  &\frac{2 \, K \, \lvert v - u \rvert  \,  \lVert A \rVert  \, \lvert B_{k/4}(z) \rvert \, f(\lvert v-u \rvert)}{(1+\max(0, (k/8)^{\frac{1}{2}} - c \,\lvert v - u \rvert))^{2\nu + 2 d}} \, \frac{1}{(1+\max(0, (\lVert x -z\rVert/4)^{\frac{1}{2}} - c \, \lvert v - u \rvert))^{2d+2}}\\
        & \leq  \sum_{x \in \Z^d  \setminus B_{k/2}(z)} \frac{2 \, K  \, \lvert v - u \rvert \,   \lVert A \rVert \, \lvert B_{k/4}(z) \rvert \, f(\lvert v-u \rvert) \, 8^{\nu+d}\, 4^{d+1} \, (1+ (c\, \lvert v - u \rvert)^{2})^{\nu+2d+1}}{(1+ k)^{\nu+d} (1+ \lVert x-y \rVert)^{d+1} } \, .
    \end{align*}
    Combining all four bounds gives the desired statement. Since for all $C>0$ the function $f$ and constant $c$ can be chosen uniformly for all $(\Phi_v)_{v\in I}$ with $\sup_{v\in I} \lVert \Phi_v \rVert_{4\nu+9d+4} < C$ and due to the choice of the constant $K$, the bound derived above is also uniform in the same way.
\end{proof}

\medskip
\begin{proposition}\label{prop: LiebRobinson}
    Let  $(\Phi_v)_{v\in I}$ be a family of interactions in $\mathcal{P}_{\infty,I}^{(0)}$ and let $(\alpha_{u,v})_{(u,v) \in I^2}$ be the family of automorphisms generated by it.  Then for each  $\nu \in \N_0$ there exists a constant $c_\nu>0$ and an increasing and at most polynomially growing function $f_\nu \colon \R_+ \to \R_+$, such that for all finite disjoint sets $X,Y \subseteq \Z^d$ and all $A \in \mA_X^+$, $B\in \mA_Y$, it holds that
    \begin{align*}
        \lVert [\, \alpha_{u,v} \, A, \, B \,] \rVert 
        \leq  \lVert A \rVert \, \lVert B \rVert \, \lvert Y \rvert \,  \frac{ f_\nu(\lvert v-u \rvert) }{ ( 1 + \max( 0, \mathrm{dist}(X,Y)^{1/2} - c_\nu \, \lvert v - u \rvert) )^\nu }  \, .
    \end{align*}
    Furthermore the dependence on $(\Phi_v)_{v\in I}$ of $f_\nu$ and $c_\nu$ is such that for each $C>0$ they can be chosen uniformly for all $(\Phi_v)_{v\in I}$ with $\sup_{v\in I} \lVert \Phi_v \rVert_{2\nu+d} < C$.
\end{proposition}
\begin{proof}
    A Lieb--Robinson bound for fermion systems with long range interactions was proven in \cite{TeufelWessel25}. Note that this construction was  done only for finite lattices, but one can easily lift the bound to $\Z^d$. To this end, denote by $(\alpha_{u,v}^k)_{(u,v)\in I^2}$ the cocycle generated by a suitable finite-volume restriction of ($\Phi)_{v\in I}$. By construction of the infinite-volume automorphism, we have that $\alpha_{u,v}^k(A)$ converges to $\alpha_{u,v}(A)$ (see \cite{Bru2017} for a detailed proof). We thus observe that
    \begin{align*}
        \lVert [ \, \alpha_{u,v} \, A, \, B \, ] \rVert \leq  \lVert [ \, \alpha_{u,v}^k \, A, \, B \, ] \rVert + 2 \, \lVert (\alpha_{u,v} - \alpha_{u,v}^k) \, A \rVert \, \lVert B \rVert.
    \end{align*}
    Since the Lieb--Robinson bound is independent of the system size and the second term vanishes in the limit $k \to \infty$, the bound also holds for the left-hand side.
    
    For any $\nu > \frac{d}{2}$ and for any $\sigma \in ((d + 1) / (2 \nu + 1), \, 1)$ Theorem 6 in \cite{TeufelWessel25} provides us with the following bound:
    \begin{align*}
        \hspace{4em}&\hspace{-4em}\lVert [\, \alpha_{u,v} \, A, \, B \, ] \rVert 
        \leq 2 \, \lVert A \rVert \, \lVert B \rVert \, \min \{ \lvert X \rvert, \, \lvert Y \rvert \} \, 
        \\
        &\times \left( \e^{c_\nu \, \lvert v - u \rvert - \mathrm{dist}(X, \, Y)^{1 - \sigma}} + C_\sigma \, (1 + \mathrm{dist} (X, \, Y) )^{- 2 \sigma \nu} \, \lvert v - u \rvert \, (1 + (c_\nu \, \lvert v - u \rvert)^{d / (1 - \sigma)}) \right) \, ,
    \end{align*}
    where  $c_\nu = \max \{ 2 \e \, \kappa_\nu \, \sup_{t \in I} \lVert \Phi_t \rVert_{2\nu}, \, \sup_{t \in I} \lVert \Phi_t \rVert_{2\nu + d}\}$, with a constants $\kappa_\nu$ and $C_\sigma$. We take $\nu > d + 1$ and choose $\sigma = \frac{1}{2}$ and consider two different cases. For $\mathrm{dist}(X, \, Y)^{1 / 2} \leq c_\nu \lvert v -u \rvert$ we use the trivial bound
    \begin{align*}
        \lVert [\, \alpha_{u,v} \, A, \, B \, ] \rVert &\leq 2 \, \lVert A \rVert \, \lVert B \rVert \, .
    \end{align*}
    For $\mathrm{dist}(X, \, Y)^{1 / 2} > c_\nu \lvert v -u \rvert$ we first notice that
    \begin{align*}
        (1 + \mathrm{dist}(X, \, Y))^\nu \geq (1 + \mathrm{dist}(X, \, Y)^{1/2} - c_\nu \, \lvert v - u \rvert )^\nu 
    \end{align*}
    and 
    \begin{align*}
        \e^{c_\nu \, \lvert v - u \rvert - \mathrm{dist}(X, \, Y)^{1 / 2}} \leq \frac{\Tilde{C}_\nu}{(1 + \mathrm{dist}(X, \, Y)^{1/2} - c_\nu \, \lvert v - u \rvert)^\nu},
    \end{align*}
    for some constant $\Tilde{C}_\nu$. Hence, we have
    \begin{align*}
        \lVert [\, \alpha_{u,v}& \, A, \, B \, ] \rVert \\ &\leq 2 \, \lVert A \rVert \, \lVert B \rVert \, \lvert Y \rvert \, \frac{1}{(1 + \mathrm{dist}(X, \, Y)^{1/2}- c_\nu \lvert v -u \rvert )^{\nu}} \,\left( \Tilde{C}_\nu + C_\sigma \, \lvert v - u \rvert \, (1 + (c_\nu \, \lvert v - u \rvert)^{2d}) \right) \, .
    \end{align*}
    Combining these two statements gives the desired bound.
\end{proof}

\medskip
\begin{lemma}\label{lem: function estimate}
    For all $\nu \in  \N_0$, $t>0$, $n \in N$ and $x \in [0,\infty)$ it holds that
    \begin{align*}
        \frac{(1+x)^\nu}{(1+ \max(0,(x/n)^\frac{1}{2} - t))^{2\nu}} \leq n^\nu (1+ t^{2})^\nu\, .
    \end{align*}
    \begin{proof}
        We observe that $$\frac{(1+x)^\nu}{(1+ \max(0,(x/n)^\frac{1}{2} - t))^{2\nu}} \leq n^\nu \frac{(1+x/n)^\nu}{(1+ \max(0,(x/n)^\frac{1}{2} - t))^{2\nu}}$$ and further that for $x\leq n\,t^{2}$  
        \begin{align*}
            n^\nu \frac{(1+x/n)^\nu}{(1+ \max(0,(x/n)^\frac{1}{2} - t))^{2\nu}} = n^\nu (1+x/n)^\nu\, ,
        \end{align*}
        which is a monotonously increasing function in $x$ on the interval $[0,n\,t^{2}]$. For $x\geq n\, t^{2}$ we have 
        \begin{align*}
            n^\nu \frac{(1+x/n)^\nu}{(1+ \max(0,(x/n)^\frac{1}{2} - t))^{2\nu}} = n^\nu \frac{(1+x/n)^\nu}{(1+ (x/n)^\frac{1}{2} - t)^{2\nu}}, 
        \end{align*}
        for which a simple analysis shows that as a function of $x$ on the interval $[n\,t^{2}, \infty)$ it attains its maximum at $x=n\,t^{2}$. Therefore, the desired bound holds.
    \end{proof}
\end{lemma}

\section{Diagonal and Off-Diagonal Maps}\label{sec: off-diag}

\begin{definition}\label{def: weight function}
    Let ($(\w_s)_{s\in I}$, $(H_s)_{s\in I})$ be a differentiable path of gapped systems and let ${g_0}$ be the supremum of the set of all $c$, such that for all $s \in I$  the state $\w_s$ is a gapped ground state of $H_s$ with gap $c$. We define $g\coloneq \min \{ g_0, 1 \}$ and the functions $w \colon \R \to \R $ and $W \colon \R \to \R $ as the functions from \cite[Lemma 2.3]{Bachmann2011}) and \cite[Lemma 2.6]{Bachmann2011}) with respect to $g$. We note the properties of $w$ and $W$ that are relevant here:
    
    The function $w$ is nonnegative, $\lVert \cdot \rVert_1$-normalized and satisfies: 
    \begin{itemize}
        \item[(i)] $w$ is even.
        \item[(ii)] The Fourier transform $\widehat w$ of $w$ is supported in $[-g,g]$.
        \item[(iii)] $\sup_{s\in\R} \lvert s \rvert^n \, w(s) < \infty$ for all $n\in \N_0$.
    \end{itemize}
    
    The function $W\colon \R \to \R$ is defined as 
    \begin{align*}
        W(t) \coloneq 
        \begin{cases} 
            \int_t^\infty \dd u \, w(u) & t \geq 0 \\
            -\int_{-\infty}^t \dd u\,  w(u) & t < 0 \, ,
        \end{cases}
    \end{align*}
    and satisfies the following properties:
    \begin{itemize}
        \item[(i)] $W$ is odd.
        \item[(ii)] The Fourier transform $\widehat W$ of $W$ satisfies $\widehat{W}(k) = \frac{- \i}{\sqrt{2  \pi}\, k}$ for  $k \in \R \setminus [-g,g]$.
        \item[(iii)] $\sup_{s\in\R} \lvert s \rvert^n \, W(s) < \infty$ for all $n\in \N_0$.
    \end{itemize}
\end{definition}

\bigskip

\begin{definition}
    Let $((\w_s)_{s\in I}, (H_s)_{s\in I})$ be a differentiable path of gapped systems and $A \in \mA$. We define the inverse Liouvillian and diagonal part of $A$ as
    \begin{align*}
        \mI_s(A) \coloneq \int_\R \dd t \,  W(t) \, \e^{\i t \mL_{H_s}} \, A\, ,
    \end{align*}
    \begin{align*}
        A\D[s] \coloneq \int_\R \dd t \, w(t) \, \e^{\i t \mL_{H_s}} \, A \, ,
    \end{align*}
    and if $A\in \mA_\infty$, we define the off-diagonal part as
    \begin{align*}
        A\OD[s] \coloneq -\i\, \mL_{H_s}  \, \mI_s(A)\, .
    \end{align*}
\end{definition}

\medskip

\begin{lemma}\label{lem: od-splitting}
    Let $((\w_s)_{s\in I}, (H_s)_{s\in I})$ be a differentiable path of gapped systems and $A\in \mA_\infty$. Then $\mI_s(A)$, $A\D[s]$, and $A\OD[s]$ lie in $\mA_\infty$ and it holds that $A = A\D[s] + A\OD[s]$.
\end{lemma}
\begin{proof}
    The first statement is a direct corollary of Lemma \ref{Lemma: Lieb-Robinson bound for D_infty} and Lemma \ref{lem: sum representation of generator}. To see the second claim, notice that $\frac{\dd}{\dd t} W(t) = - w (t)$ and hence integration by parts gives us that 
    \begin{align*}
        \int_\R \dd t \, w(t) \, \int_0^t \dd u \, \e^{\i u \mL_{H_s}} \, A \;=\; - W(t) \int_0^t \dd u \, \e^{\i u \mL_{H_s}} \, A \, \biggr \rvert_{-\infty} ^\infty + \int_\R \dd t \,  W(t) \, \e^{\i u \mL_{H_s}} \, A \, ,
    \end{align*}
    where the boundary terms vanish due to the fall-off property of $W$. Then we can calculate
    \begin{align*}
        A\D[s] + A\OD[s] &= \int_\R \dd t \, w(t) \, (\e^{\i t \mL_{H_s}} - \i \, \int_0^t \dd u \, \mL_{H_s} \, \e^{\i u \mL_{H_s}} ) \, A\\
        &= \int_\R \dd t \, w(t) \, (\e^{\i t \mL_{H_s}} - \e^{\i t \mL_{H_s}} + 1) \, A\\
        &= A \, ,
    \end{align*}
    where we have used that $w$ is normalised.
\end{proof}

\medskip

\begin{definition}
    Let $((\w_s)_{s\in I}, (H_s)_{s\in I})$ be a differentiable path of gapped systems and  $\Phi \in \mathcal{P}_\infty$. We define the inverse Liouvillian of $\Phi$ in two steps: First we define, for $x \in \Z^d$, the quasi-local observables 
    \begin{align*}
        \mI_s(\Phi)_{x,*} \coloneq -\i \int_\R \dd t \, W(t)  \int_0^t \dd u \, \e^{\i u \mL_{H_s}} \,  \mL_\Phi \, H_{s,x} \, .
    \end{align*}
    We then define the interaction $\mI(\Phi)$ via
    \begin{align*}
        \mI_s(\Phi)(B_k(x)) & = (\E_{B_k(x)} - \E_{B_{k-1}(x)}) \, \mI_s(\Phi)_{x,*}\\
        \mI_s(\Phi)(B_0(x)) & = \E_{B_0(x)} \, \mI_s(\Phi)_{x,*} \, ,
    \end{align*}    
    for all $(x,k) \in \Z^d \times \N$ and as $\mI_s(\Psi)(M)= 0$ for all $M \in P_0(\Z^d)$ that are not equal to $B_k(x)$ for some $(x,k) \in \Z^d \times \N_0$.
\end{definition}

\medskip
\begin{remark}
    Note that the inverse Liouvillian of an interaction is defined such that for all $x \in \Z^d$ it holds that $\mI_s(\Phi)_x=\mI_s(\Phi)_{x,*}$
\end{remark}

\medskip

\begin{lemma} \label{Lemma: inverse Liouvillian: B->B}
    Given a differentiable path of gapped systems $((\w_s)_{s\in I}, (H_s)_{s\in I})$, it holds for any $(\Phi_s)_{s\in I} \in \mathcal{P}_{\infty,I}^{(0)}$ 
    that $(\mI_s(\Phi_s))_{s\in I}\in B^{(0)}_{\infty, I}$. In particular $(\mI_s(\Phi_s))_{s\in I}$ is the generator of a unique cocycle of automorphisms (by Lemma \ref{lem: cocycles}).
\end{lemma}
\begin{proof}
    We first note that for all $\nu \in \N_0$ we have $\sup_{x \in \Z^d }\lVert H_{s,x} \rVert_{\nu,x} \leq 3 \, \lVert H_s \rVert_\nu$. Hence, we find by Lemma \ref{Lemma: Lieb-Robinson bound for D_infty} and Lemma \ref{lem: sum representation of generator} that $\sup_{s \in I} \sup_{ x \in \Z^d} \lVert \mI_s (\Phi_s)_x \rVert_{\nu , x} < \infty$.\\
    Now we can calculate
    \begin{align*}
        \lVert \mI_s( \Phi_s ) \rVert_\nu &= \sup_{x \in \Z^d} \, \sum_{ \substack{M \in P_0(\Z^d) \\ x \in M }} (1 + \diam (M) )^\nu \, \lVert \mI_s(\Phi_s)(M) \rVert\\
        &\leq \sup_{x \in \Z^d} \, (  \sum_{ k = 1 }^\infty \, \sum_{ y \in B_k(x)} (1 + 2k )^\nu \, \lVert (\E_{ B_k ( y ) } - \E_{B_{k-1}( x ) } ) \, \mI_s ( \Phi_s )_y \rVert + \lVert \E_{B_0(x)} \, \mI_s ( \Phi_s )_x \rVert) \\
        &\leq \sum_{ k = 1 }^\infty \frac{(1 + 2 k )^d \, (1 + 2k )^\nu}{(1 + (k - 1) )^{\nu + d + 2}} \,  \sup_{x \in \Z^d} \, \lVert  \mI_s ( \Phi_s )_x \rVert_{\nu + d + 2, x} + \sup_{x \in \Z^d} \, \lVert \mI_s ( \Phi_s )_x \rVert \, ,    
    \end{align*}
    which gives $\sup_{s\in I}\lVert \mI_s(\Phi_s) \rVert_\nu < \infty$ for all $\nu \in \N_0$.

    To see the continuity, we notice that
    \begin{align*}
        \lVert \mI_{s+h} (\Phi_{s+h}) (M) - \mI_s (\Phi_s) (M) \rVert \leq 2 \, \lVert \mI_{s+h} (\Phi_{s+h})_x - \mI_s (\Phi_s)_x \rVert \, ,
    \end{align*}
    for some $x \in \Z^d$. Now compute   
    \begin{align*}
       &\lVert \mI_{s+h} (\Phi_{s+h})_x - \mI_{s} (\Phi_{s})_x \rVert \\
        & \quad \leq \int_\R \dd t \, W(t) \int_0^t \dd u \, \lVert \e^{\i u \mL_{H_{s+h}}} \mL_{ \Phi_{s+h} } \, H_{s+h,x} - \e^{\i u \mL_{H_s}} \mL_{ \Phi_{s} } \, H_{s,x} \rVert \, .   
    \end{align*}
    We split the integrand as follows
    \begin{align*}
        \lVert \e^{\i u \mL_{H_{s+h}}} \mL_{ \Phi_{s+h} } \, H_{s+h,x} - \e^{\i u \mL_{H_s}} \mL_{ \Phi_{s} } \, H_{s,x} \rVert 
        & \leq \lVert \mL_{\Phi_{s+h}}\, (H_{s+h,x} - H_{s,x} ) \rVert \\
        &\quad + \lVert  (\mL_{\Phi_{s+h}} - \mL_{\Phi_s}) \, H_{s,x} \rVert \\
        & \quad + \lVert ( \e^{\i u \mL_{H_{s+h}}} - \e^{\i u \mL_{H_s}}) \, \mL_{\Phi_s} \, H_{s,x} \rVert \, .
    \end{align*}
    The first term vanishes in the limit $h \to 0$ due to Lemmas \ref{lem: sum representation of generator} and \ref{lem: derivative of zero chain}. The second term vanishes in the limit $h \to 0$ by Lemma \ref{lem: derivative of Liouvillian}. For the last term we observe that by Duhamel's formula
    \begin{align*}
        \lVert (\e^{ \i u \mL_{H_{s + h}}} - \e^{\i u \mL_{H_s}}) \, \mL_{\Phi_s} \, H_{s,x} \rVert &\leq \int_0^h \d t \, \lVert \partial_t \, \e^{\i u \mL_{H_{s + t}}} \, \mL_{\Phi_s} \, H_{s,x} \rVert
        \\
        & \leq u \, \int_0^h \d t \,  \int_0^u \d v \, \lVert \e^{\i v \mL_{H_{s + t}}} \, \mL_{\Dot{H}_{s+t}} \, \e^{\i (u - v) \mL_{H_{s + t}}} \, \mL_{\Phi_s} \, H_{s,x} \rVert \, .
    \end{align*}    
    This term also vanishes in the limit $h \to 0$ by Lemma \ref{lem: sum representation of generator} and Lemma \ref{Lemma: Lieb-Robinson bound for D_infty}. The claim then follows by dominated convergence.
\end{proof}

\medskip
\begin{lemma}\label{lem: Inverse Liouvillian on derivations}
    Let $((\w_s)_{s\in I}, (H_s)_{s\in I})$ be a differentiable path of gapped systems, $\Phi\in \mathcal{P}_\infty$, and $A\in \mA_\infty$. For all $s\in I$ it holds that
    \begin{align*}
        \int_\R \dd t \,  W(t)\, \e^{\i t \mL_{H_s}} \, \mL_{\Phi} \, \e^{-\i t \mL_{H_s}}\, A = \mL_{\mI_s(\Phi)}\,A \, .
    \end{align*}
\end{lemma}
\begin{proof}
    We first notice that by Lemma \ref{Lemma: inverse Liouvillian: B->B} it is $\mI_s(\Phi)\in \mathcal{P}_\infty$, and hence by Lemma \ref{lem: sum representation of generator} we find
    \begin{align*}
        \mL_{\mI_s ( \Phi )} \, A &= \sum_{x \in \Z^d} \, [ \mI_s ( \Phi )_x , \, A ]\\
        &= - \sum_{x\in\Z^d} \, \i \int_{\R} \dd t\, W(t)\,\int_0^t\dd u \, [ \e^{ \i u \mL_{H_s}} \, \mL_{\Phi} \, H_{s,x} , \, A ]\\
        &= - \sum_{x\in\Z^d} \, \i \int_{\R} \dd t\, W(t)\,\int_0^t\dd u \, [ \e^{\i u \mL_{H_s} } \, \sum_{ y \in \Z^d} \, [\Phi_y , \, H_{s,x}] , \, A ] \, ,
    \end{align*}
    where the last equality holds since $H_{s,x}\in \mA_\infty$. Using Lemma \ref{Lemma: Lieb-Robinson bound for D_infty} and Lemma \ref{lem:commutator bound}, we get absolute summability/integrability, and hence we can exchange the order of summation.
    \begin{align*}
        \mL_{\mI_s ( \Phi )} \, A &= \sum_{y\in\Z^d} \, [ \i \int_{\R} \dd t\, W(t) \, \int_0^t \dd u \, \e^{ \i u \mL_{H_s} } \, \mL_{H_s} \, \Phi_y , \, A ]\\
        &= \sum_{y\in\Z^d} \, [  \int_{\R} \dd t\, W(t) \, \e^{\i t \mL_{H_s}} \, \Phi_y , \, A ]\\
        &= \int_{\R} \dd t\, W(t) \, \e^{\i t \mL_{H_s}} \, \sum_{y\in\Z^d} \, [ \Phi_y , \, \e^{-\i t \mL_{H_s}} \, A ]\\
        &= \int_\R \dd t \,  W(t)\, \e^{\i t \mL_{H_s}} \, \mL_{\Phi} \, \e^{-\i t \mL_{H_s}}\, A \, ,
    \end{align*}
    where we used that $W(t)$ is an odd function in the second equality.
\end{proof}

\section{Technical Lemmas}

\medskip
\begin{lemma} \label{lem: derivative of zero chain}
    Let $I \subseteq \R$ be an interval, $(\Phi_s)_{s \in I} \in \mathcal{P}_{\infty, I}^{(k)}$ and $x \in \Z^d$. The map $s \mapsto \Phi_{s, x}$ is $k$ times continuously differentiable with respect to $\lVert \cdot \rVert _{\nu, y}$ for all $\nu \in \N_0$ and $y \in \Z^d$. For $0\leq m \leq k$ the derivative is given by $\frac{\d^m}{\d s^m} (\Phi_{s,x})  =(\frac{\d^m}{\d s^m} \Phi_s)_x$, where $\frac{\d^m}{\d s^m}\Phi_s$ denotes the interaction obtained by term-wise differentiation.
\end{lemma}
\begin{proof}
    We show that if $(\Phi_s)_{s \in I} \in \mathcal{P}_{\infty, I}^{(0)}$ then $s \mapsto \Phi_{s, x}$ is continuous with respect to $\lVert \cdot \rVert _{\nu, y}$ for all $\nu \in \N_0$ and $y \in \Z^d$ and that if $(\Phi_s)_{s \in I} \in \mathcal{P}_{\infty, I}^{(1)}$ then $s \mapsto \Phi_{s, x}$ is differentiable with respect to $\lVert \cdot \rVert _{\nu, y}$ for all $\nu \in \N_0$ and $y \in \Z^d$ with $ \Dot{(\Phi_{s,x})} = (\Dot{\Phi}_{s})_x$. The full statement then follows inductively.
    
    So for the first part let $(\Phi_s)_{s \in I} \in \mathcal{P}_{\infty, I}^{(0)}$. We first notice that
    \begin{align*}
        \Phi_{s+h, x} - \Phi_{s, x} = \sum_{M \in R_x} \Phi_{s+h}(M) - \Phi_s (M) = \sum_{j = 0}^\infty \sum_{\substack{M \in R_x \\ \mathrm{diam}(M) = j}} \Phi_{s+h}(M) - \Phi_s (M) \, ,
    \end{align*}
    where the inner sum has only finitely many summands. We now look at the the sum
    \begin{align*}
        \sum_{\substack{M \in R_x \\ \mathrm{diam}(M) = j}} \lVert \Phi_s (M) \rVert_{\nu, y}
        &\leq \sum_{\substack{M \in R_x \\ \mathrm{diam}(M) = j}} (1+ \lVert x-y \rVert)^\nu \, \lVert \Phi_s (M) \rVert_{\nu, x}
        \\
        &\leq \frac{3\, (1+ \lVert x-y \rVert)^\nu}{(1 + j)^2} \sum_{\substack{M \in R_x \\ \mathrm{diam}(M) = j}} (1 + j)^{\nu + 2} \, \lVert \Phi_s (M) \rVert
        \\
        &\leq \frac{3\, (1+ \lVert x-y \rVert)^\nu}{(1 + j)^2} \, \lVert \Phi_s \rVert_{\nu + 2} \, ,
    \end{align*}
    and therefore find that
    \begin{align*}
        \lVert \sum_{\substack{M \in R_x \\ \mathrm{diam}(M) = j}} \Phi_{s+h}(M) - \Phi_s (M) \rVert_{\nu, y} 
        \leq  \frac{6\, (1+ \lVert x-y \rVert)^\nu}{(1 + j)^2} \, \sup_{s \in I} \lVert \Phi_s \rVert_{\nu + 2}
        \, .
    \end{align*}
    This expression is independent of $h$ and summable. Therefore the dominated convergence theorem applies and we find that
    \begin{align*}
        \lim_{h \to 0} \, \lVert \Phi_{s+h, x} - \Phi_{s, x} \rVert_{\nu, y} &\leq \sum_{j = 0}^\infty \sum_{\substack{M \in R_x \\ \mathrm{diam}(M) = j}} \lim_{h \to 0} \, \lVert \Phi_{s+h} (M) - \Phi_s (M) \rVert_{\nu, y}
        \\
        & \leq 3 \sum_{j = 0}^\infty \sum_{\substack{M \in R_x \\ \mathrm{diam}(M) = j}} (1+ \lVert x-y \rVert)^\nu \, (1 + j)^\nu \, \lim_{h \to 0} \, \lVert \Phi_{s+h} (M) - \Phi_s (M) \rVert 
        \\
        &= 0 \, ,
    \end{align*}
    this shows the continuity of $s \mapsto \Phi_{s, x}$. 
    
    Now let $(\Phi_s)_{s \in I} \in \mathcal{P}_{\infty, I}^{(1)}$, to see the the differentiability, note that for all $M \in P_0(\Z^d)$
    \[
    \Phi_{s+h} (M) - \Phi_{s} (M) = \int_{s}^{s+h} \Phi^\prime_{t} (M) \, \d t 
    \]
    and that thus, with the same argument as before
    \begin{align*}
        \lVert \sum_{\substack{M \in R_x \\ \mathrm{diam}(M) = j}} \frac{\Phi_{s+h}(M) - \Phi_s (M)}{h} - \Dot{\Phi}_s(M) \rVert_{\nu, y} \leq  \frac{6\, (1+ \lVert x-y \rVert)^\nu}{(1 + j)^2} \, \sup_{s \in I} \lVert \Dot{\Phi}_s \rVert_{\nu + 2} \, .
    \end{align*}
    So again, by dominated convergence we can pull the limit through the sum and conclude the differentiability and that $ \Dot{(\Phi_{s,x})} = (\Dot{\Phi}_{s})_x$.
\end{proof}

\medskip
\begin{lemma}\label{lem: derivative of Liouvillian}
Let $I \subseteq \R$ be some interval, $(\Phi_s)_{s \in I} \in \mathcal{P}_{\infty, I}^{(k)}$ and $A \in \mA_\infty$. The map $s \mapsto \mL_{\Phi_s} \, A$ then is $k$ times continuously differentiable with respect to $\lVert \, \cdot \, \rVert_{\nu, x}$ for all $x \in \Z^d$ and $\nu \in \N_0$. For all $0 \leq m \leq k$ the derivative is given by $\frac{\d^m}{\d s^m} (\mL_{\Phi_s}\, A) = \mL_{\frac{\d^m}{\d s^m}\Phi_s} \, A$, where $\frac{\d^m}{\d s^m}\Phi_s$ denotes the interaction obtained by term-wise differentiation.
\end{lemma}
\begin{proof}
    We show that if $(\Phi_s)_{s \in I} \in \mathcal{P}_{\infty, I}^{(0)}$ then $s \mapsto \mL_{\Phi_s} \, A$ is continuous with respect to $\lVert \cdot \rVert _{\nu, x}$ for all $\nu \in \N_0$ and $x \in \Z^d$ and that if $(\Phi_s)_{s \in I} \in \mathcal{P}_{\infty, I}^{(1)}$ then $s \mapsto \mL_{\Phi_s} \, A$ is differentiable with respect to $\lVert \cdot \rVert _{\nu, x}$ for all $\nu \in \N_0$ and $x \in \Z^d$ with $ \Dot{(\mL_{\Phi_s}\, A)} = \mL_{\Dot{\Phi}_s} \, A$. The full statement then follows inductively.

    So for the first part let $(\Phi_s)_{s \in I} \in \mathcal{P}_{\infty, I}^{(0)}$. We can use Lemma \ref{lem:commutator bound} to obtain
    \begin{align*}
        \lVert \mL_{\Phi_{s+h}} \, A - \mL_{\Phi_s} \, A \rVert_{\nu, x} 
        &\leq \sum_{y \in \Z^d} \, \lVert [ (\Phi_{s + h, y} - \Phi_{s, y}), \, A] \rVert_{\nu, x}
        \\
        &\leq \sum_{y \in \Z^d} 2 \, \sup_{s \in I} 4^{\nu + d + 4} \, \frac{\lVert \Phi_{s, y} \rVert_{\nu + d + 1, y} \, \lVert A \rVert_{\nu + d + 1, x}}{(1 + \lVert x - y \rVert)^{d + 1}}
        \\
        &\leq 2 \,  \sup_{s \in I} \sup_{z \in \Z^d} \lVert \Phi_{s, z} \rVert_{\nu + d + 1, z} \, \lVert A \rVert_{\nu + d + 1, x} \sum_{y \in \Z^d} \frac{4^{\nu + d + 4}}{(1 + \lVert x - y \rVert)^{d + 1}} 
        \\
        &\leq 6 \, \sup_{s \in I}  \lVert \Phi_{s} \rVert_{\nu + d + 1} \lVert A \rVert_{\nu + d + 1, x} \, \sum_{y \in \Z^d} \frac{4^{\nu + d + 4}}{(1 + \lVert x - y \rVert)^{d + 1}}
        \, .
    \end{align*}
    Hence, we can pull the limit $h\to 0$ inside the sum by dominated convergence. Since the map $s \mapsto \Phi_{s, x}$ is continuous by Lemma \ref{lem: derivative of zero chain} we get the continuity of $s \mapsto \mL_{\Phi_s} \, A$. 
    
    Now let $(\Phi_s)_{s \in I} \in \mathcal{P}_{\infty, I}^{(1)}$. By Lemma \ref{lem: derivative of zero chain} the map $ s\mapsto \Phi_{s,y}$ is continuously differentiable for each $y\in \Z^d$, giving us
    \begin{align*}
        \Phi_{s+h, y} - \Phi_{s, y} = \int_{s}^{s+h} \Phi^\prime_{t, y} \, \d t \, .
    \end{align*}
    Thus we find as before
    \begin{align*}
        \lVert \frac{\mL_{\Phi_{s+h}} \, A - \mL_{\Phi_s} \, A}{h} - \mL_{\Dot{\Phi}_s} \, A\rVert_{\nu, x} 
        &\leq \sum_{y \in \Z^d} \, \lVert [ (\frac{\Phi_{s + h, y} - \Phi_{s, y}}{h} - \Dot{\Phi}_{s, y}), \, A] \rVert_{\nu, x}
        \\
        &\leq 6 \, \sup_{s \in I}  \lVert \Dot{\Phi}_{s} \rVert_{\nu + d + 1} \lVert A \rVert_{\nu + d + 1, x} \, \sum_{y \in \Z^d} \frac{4^{\nu + d + 4}}{(1 + \lVert x - y \rVert)^{d + 1}}
        \, ,
    \end{align*}
    allowing us to pull the limit $h\to 0$ into the sum, which results in the differentiability of $s \mapsto \mL_{\Phi_s} \, A$ and $ \Dot{(\mL_{\Phi_s} \, A)} = \mL_{\Dot{\Phi}_s}\, A $.
\end{proof}

\medskip
\begin{lemma}\label{lem: derivative switch}
    Let $((\w_s)_{s\in I}, (H_s)_{s\in I})$ be a differentiable path of gapped systems and $A\in \mA_\infty$. For all $s\in I$ it holds that 
    \begin{align*}
        \Dot{\w}_s(\mL_{H_s}A)\, =\, - \w_s(\mL_{\Dot{H}_s} A) \, .
    \end{align*}
\end{lemma}
\begin{proof}
    Since $\w_s$ is an $\mL_{H_s}$-ground state the following holds
    \begin{align*}
        \Big| \Dot{ \w }_s ( \mL_{ H_s } &\left. \, A ) + \w_s ( \mL_{ \Dot{H}_s } \, A ) \right| = \left| \Dot{ \w }_s ( \mL_{H_s} \, A ) + \w_s ( \mL_{ \Dot{H}_s } \, A ) - \frac{ \w_{s + h} ( \mL_{ H_{s+h}} \, A ) - \w_{s} ( \mL_{ H_s } \, A ) }{h} \right|
        \\
        &\leq \left|\w_{s+h} \left( \, \, 
        \frac{ \mL_{ H_{s+h}} \, A - \mL_{H_s} \, A }{h} - \mL_{\Dot{H}_s} \, A \right) \right| + \left| \frac{ \w_{s+h} ( \mL_{H_s} \, A) - \w_{s} ( \mL_{H_s} \, A ) }{h} -\Dot{\w}_s ( \mL_{H_s} \, A ) \right|
        \\
        & \quad +\left| (\w_{s+h} - \w_s ) ( \mL_{ \Dot{H}_s } \, A ) \right| \, .
    \end{align*}
    The last two terms vanish by the continuity and differentiability of $\w_s$ in the limit $h \to 0$. For the other remaining term, we use that $\w_{s+h}$ is a state and hence
    \begin{align*}
        \left|\w_{s+h} \left(\frac{ \mL_{ H_{s+h} } \, A - \mL_{ H_s } \, A}{ h } - \mL_{ \Dot{H}_s } \, A \right) \right| \leq \left\lVert \mL_{ \frac{ H_{s+h} - H_s }{ h } - \Dot{H}_s} \, A \right\rVert \, .
    \end{align*}
    This term vanishes in the limit $h \to 0$ by Lemma \ref{lem: derivative of Liouvillian}.
\end{proof}

\medskip
\begin{lemma} \label{Lemma: Lieb-Robinson bound for D_infty}
    Let $((\w_s)_{s\in I}, (H_s)_{s\in I})$ be a differentiable path of gapped systems and $A\in \mA_\infty$. For each $\nu\in \N_0$ there exists an increasing and at most polynomially growing function  function $ b_\nu \colon [0,\infty) \to [0,\infty)$, such that for all $t\in \R$ and $z \in \Z^d$, it holds that
    \begin{align*}
        \sup_{s\in I} \left\lVert \e^{\i t \mL_{H_s}} \, A \right\rVert_{\nu,z} 
        \leq b_\nu(\lvert t \rvert) \, \lVert A \rVert_{\nu,z} \, .
    \end{align*}
\end{lemma}

\begin{proof}
    By Assumption \ref{Assumption: Hs is in B infty} the $H_s$ satisfies the conditions of Lemma \ref{lem: cocycles} for each $s\in I$. Since there exists a $C > 0$ such that $\sup_{s \in I} \lVert H_s \rVert_{ 4\nu + 9d + 4} < C$, we can choose the function $b_\nu$ uniformly for all $s \in I$. Then we have that
    \[
        \sup_{s\in I} \left\lVert \e^{\i t \mL_{H_s}} \, A \right\rVert_{\nu,z} 
        \leq b_\nu(\lvert t \rvert) \, \lVert A \rVert_{\nu,z} \, .\qedhere
    \]
\end{proof}

\medskip
\textbf{Acknowledgments}. We thank Tom Wessel for helpful discussions.
This work was supported by the Deutsche Forschungsgemeinschaft (DFG, German Research Foundation) through TRR 352 (470903074) and FOR 5413 (465199066).

\printbibliography

\end{document}